\documentstyle[preprint,aps]{revtex}
\input epsf

\begin{document}
\newcommand{\none}{\nonumber\\}
\newcommand{\bea}{\begin{eqnarray}}
\newcommand{\eea}{\end{eqnarray}}
\draft


\tighten

\title{ Two-loop Quantum Corrections of Scalar QED with Non-minimal 
 Chern-Simons Coupling}

\author{ M.E. Carrington${}^{a,c}$, W.F. Chen${}^{b,c}$, G. Kunstatter${}^{b,c}$ and J. Mottershead${}^{b,d}$}
 
\address{ ${}^a$ Department of Physics, Brandon University, Brandon, Manitoba, R7A 6A9 Canada\\
 ${}^b$ Department of Physics, 
 University of Winnipeg, Winnipeg, Manitoba, R3B 2E9 Canada \\
${}^c$  Winnipeg Institute for
Theoretical Physics, Winnipeg, Manitoba\\
${}^d$ {\it Permanent Address:} Physics Department, University of Alberta
   Edmonton, Alberta T6G 2J1, Canada}

\maketitle
\begin{abstract}
We investigate two-loop 
quantum corrections to  
non-minimally coupled Maxwell-Chern-Simons theory. The non-minimal 
gauge interaction represents the  magnetic moment interaction 
between the charged scalar and the electromagnetic field. We show that
one-loop renormalizability of the theory found in previous work does 
not survive to two-loop level. 
However, with an appropriate 
choice of the non-minimal coupling constant, 
it is possible to 
renormalize  the two-loop effective potential and hence render it potentially 
useful for a detailed analysis of 
spontaneous symmetry breaking induced by radiative corrections.
\end{abstract}

\vspace{3ex}

\section{Introduction}

Maxwell-Chern-Simons electrodynamics has been studied extensively in 
recent years for a variety of reasons. The Chern-Simons term gives
 the photon a topological mass without spontaneously breaking gauge 
 symmetry$\cite{refn1}$ 
and allows for the existence of charged particles with
fractional statistics$\cite{refn2}$.  
Pure Chern-Simons scalar electrodynamics 
admits topological and non-topological self-dual solitons, 
for which many exact solutions to the classical equations 
of motion are available$\cite{refn3}$. 
 Moreover, such theories may also have physical significance.  
Relativistic three dimensional Chern-Simons theories
provide a consistent  description of 
the high temperature limit of four dimensional gauge theories$\cite{refn4}$, 
and certain solid state systems with planar dynamics$\cite{refn2}$.
In addition, the non-relativistic version of Maxwell-Chern-Simons theory has 
been applied to the fractional Hall effect, and more 
recently to rotating superfluid ${}^3$He-A\cite{goryo}.

Recently a version of scalar electrodynamics in three dimensions 
has been studied in which
a non-minimal Chern-Simons type gauge interaction was introduced.
The non-minimal coupling in this model represents a
magnetic moment interaction between the charged scalar 
and the electromagnetic field. It is of interest for several reasons. Firstly,
it is well known that one of the most important features of scalar 
quantum electrodynamics (QED) is 
the occurrence of the Coleman-Weinberg mechanism$\cite{refn5}$. In scalar QED 
with non-minimal coupling, the Chern-Simons term is generated 
through the Coleman-Weinberg mechanism$\cite{ref1}$.  In this sense, 
the non-minimal model is the one in which the Chern-Simons term arises 
naturally 
rather than being put in by hand.  

Another reason that the non-minimal model is of interest involves 
the study of vortex solutions.  In recent years, the classical
vortex solutions of 2+1-dimensional
Chern-Simons field theories have received considerable
 attention$\cite{refn3,ref3}$. To find such a solution exactly, 
the model must be self-dual. A self-dual theory is one in which 
the classical equations of motion can be reduced from second- to 
first-order differential equations. 
  In the absence of a Maxwell term, scalar
QED with a Chern-Simons term is self-dual, and the 
topological and non-topological
vortex solutions have been found with an appropriately
chosen scalar potential$\cite{refn3}$. However, 
if the Maxwell term is present, a self-dual
 Maxwell-Chern-Simons theory can be achieved only if a magnetic 
moment interaction between the scalar and the gauge field,
i.e. the non-minimal Chern-Simons coupling, 
is introduced$\cite{ref3,ref5}$. 

It is well known that 
Maxwell-Chern-Simons scalar QED is renormalizable. 
Non-minimal gauge interactions are, however, 
notoriously non-renormalizable in four dimensions. There is
some hope that the situation might
be different in three dimensions. Some time ago it was found by 
two of us that the non-minimal
Chern-Simons coupling in 2+1-dimensional scalar electrodynamics 
is actually renormalizable at the one-loop level$\cite{ref1}$.
 The renormalizability occurs because the non-minimal gauge interaction
contains 
the three-dimensional antisymmetric tensor.  An analysis \cite{ref1} of 
the symmetry breaking by induced radiative corrections  
shows  that at the one-loop level
the non-minimal model behaves differently from the minimally coupled one. 
 In the usual Maxwell-Chern-Simons scalar QED,  
symmetry breaking results from quantum corrections but depends 
on the choice of a renormalization scale$\cite{ref6}$,
 whereas in the non-minimal model the symmetry breaking is 
unambiguous and there is a finite temperature phase transition to 
the symmetry restored state.

It is clearly of interest to discover whether the renormalizability 
of this model persists beyond one-loop level and to 
 compare the symmetry breaking phase transitions in the minimal 
and non-minimal models at higher orders. 
The two-loop behaviour 
of the minimal model has recently been analyzed in detail$\cite{ref6}$ 
where it was shown that the Coleman-Weinberg mechanism occurs at two-loops. 
The purpose of the  present paper is to commence 
 a detailed analysis of the two-loop behaviour of the non-minimally
coupled Maxwell-Chern-Simons theory. We will show
 that the model, not surprisingly,
is not renormalizable at two-loop level. However, the two-loop effective
potential can be made renormalizable providing certain conditions are
satisfied by the coupling constants 
of  the model. 
Thus we will show that under certain circumstances
(i.e. when the lowest order in the momentum expansion is sufficient) the model
may yield physically relevant predictions for the spontaneous symmetry breaking
by radiative corrections at the two-loop level.  
A detailed analysis on 
the renormalized effective potential and the consequent symmetry 
breaking is deferred to a future publication.

The paper is organized as follows.  Sect II is a brief review, 
containing an introduction to the model, a discussion of some technical 
aspects of dimensional regularization in 2+1 dimensions, and a list
of some of the necessary Feynman rules. Sect III demonstrates  
that the full effective action is not renormalizable at the two loop-level.
 Sect IV is devoted to the (somewhat lengthy) calculation of the
two-loop effective potential in the $\overline{MS}$ substraction scheme. 
In contrast to the one-loop case, the  renormalizability of the two-loop
effective potential requires a specific choice of the non-minimal 
coupling constant.  Our conclusions are summarized in Sect V, while 
some useful formulae are  collected 
in the appendix. 
 
\section{Scalar QED with non-minimal Chern-Simons Coupling}

\subsection{The Lagrangian}

The Lagrangian for scalar QED in $2+1$ dimensions with a non-minimal Chern-Simons coupling is$\cite{ref1}$
\begin{eqnarray}
{\cal L}=\frac{1}{2}(D_{\mu}\phi)^*D^{\mu}\phi-\frac{1}{4}F_{\mu\nu}F^{\mu\nu}
-\frac{i}{8}\gamma_0\epsilon^{\mu\nu\rho}F_{\nu\rho}\left[\phi^*D_{\mu}\phi-
(D_{\mu}\phi)^*\phi\right] - \frac{\lambda}{6!}(\phi^*\phi)^3,
\label{eq1}
\end{eqnarray}
where the complex scalar field $\phi$ can be decomposed into the real
and imaginary parts, $\phi=\chi+i\eta$; 
 $F_{\mu\nu}=\partial_{\mu}A_{\nu}-\partial_{\nu}A_{\mu}$;
$D_{\mu}\phi=\partial_{\mu}\phi+ieA_{\mu}\phi$, 
and $\gamma_0$, $e$ and $\lambda$
are the non-minimal Chern-Simons coupling, the gauge coupling and 
the scalar self-interaction coupling constants, respectively. The dimensional
assignments for the fields and coupling constants are as follows:
\begin{eqnarray}
[A_{\mu}]=[\phi]=[e]=M^{1/2}, ~~~~[\gamma_0]=M^{-1/2}, ~~~~[\lambda]=M^0.\label{eq1b}
\end{eqnarray}
The negative mass dimension of $\gamma_0$ indicates that the theory 
is not renormalizable in general.  

Following the standard technique to calculate the effective potential 
we first assume the existence of a non-vanishing vacuum expectation value 
for the scalar field $\langle \phi\rangle=v$ with $v$ real, and shift 
the real part of the scalar field, $\chi\rightarrow\chi +v$$\cite{ref7}$. We look for a non-vanishing value
of $v$ by determining the minimum of the effective potential 
generated by quantum corrections. Note that in the Lagrangian (\ref{eq1}),
we have put the bare mass of the scalar field, the quartic scalar
 self-interaction coupling and the statistical parameter
of a possible Chern-Simons term equal to zero, even though these terms 
are allowed by the gauge symmetry. The corresponding counterterms may 
appear, however, in the counterterm Lagrangian.   
The physical parameters can be obtained in the usual way, i.e. by 
choosing renormalization conditions that give zero renormalized 
parameters at $v=0$. 
To avoid the infrared divergences we  use 
the Landau-type $R_\xi$ gauge$\cite{ref8}$.  The gauge-fixing term is
\begin{eqnarray}
{\cal L}_{\rm g.f.}=-\frac{1}{2\xi}(\partial_{\mu}A^{\mu}-\xi ev\eta)^2, ~~~~
\xi=0.
\end{eqnarray} 
Up to a total derivative term, the Lagrangian can be divided into the
kinetic part   
and ten interaction terms:
\begin{eqnarray}
{\cal L}^{(0)} &=& -\frac{1}{2}\chi \left(\partial^2+\frac{\lambda}{4!}v^4\right)\chi
-\frac{1}{2}\eta\left(\partial^2+\frac{\lambda}{5!}v^4
-\xi e^2v^2\right)\eta \,\nonumber\\
&& +\frac{1}{2}A^{\mu}\left[(\partial^2+e^2v^2)g_{\mu\nu} -\left(1-\frac{1}{\xi}\right)\partial_{\mu}\partial_{\nu}
-\gamma_0ev^2\epsilon_{\mu\nu\rho}
\partial^{\rho}\right]A^{\nu} \, ,\nonumber\\
{\cal L}^{(1)} &=&-\lambda \left(\frac{1}{6!}\chi^6+\frac{1}{5{\times}4!{\times}2!}\chi^4\eta^2+
\frac{1}{5{\times}4!{\times}2!}\chi^2\eta^4+\frac{1}{6!}\eta^6\right)
\, ,\nonumber\\
{\cal L}^{(2)}&=&-\lambda v\left(\frac{1}{5!}\chi^5+\frac{1}{5{\times}3!{\times}2!}\chi^3\eta^2+
\frac{1}{5{\times}4!}\chi\eta^4\right) \, ,\nonumber\\
{\cal L}^{(3)}&=&-\lambda v^2\left(\frac{1}{2{\times}4!}\chi^4+
\frac{1}{10{\times}2!{\times}2!}\chi^2\eta^2+
\frac{1}{10{\times}4!}\eta^4\right) \, ,\nonumber\\
{\cal L}^{(4)}&=&-\frac{1}{3!{\times}3!}\lambda
v^3\chi^3-\frac{1}{5!{\times}3!{\times}2!}\lambda v^3\chi\eta^2 \, ,
\nonumber\\
{\cal L}^{(5)}&=&eA^{\mu}\left(\partial_{\mu}\eta \chi
-\partial_{\mu}\chi\eta\right) \, ,\nonumber \\
{\cal L}^{(6)}&=&\frac{1}{2}\gamma_0 \epsilon^{\mu\nu\rho}
\partial_\nu A_{\rho}\left(\chi\partial_\mu\eta-\eta\partial_\mu\chi\right)
\, ,\nonumber\\
{\cal L}^{(7)}&=&\frac{1}{2}e^2A_{\mu}A^{\mu}
\left(\chi^2+\eta^2\right)\, ,\nonumber\\
{\cal L}^{(8)}&=& -\frac{1}{2}\gamma_0 e\epsilon^{\mu\nu\rho}A_{\mu}\partial_{\rho}A_{\nu}
\left(\chi^2+\eta^2\right) \, ,\nonumber\\
{\cal L}^{(9)}&=&e^2v A_{\mu}A^{\mu}\chi \, ,\nonumber \\
{\cal L}^{(10)}&=&-\gamma_0 ve \epsilon^{\mu\nu\rho}
A_\mu\partial_\rho A_{\nu}\chi \, . \label{eq5}
\label{eqla}
\end{eqnarray}

\subsection{Regularization}

A regularization scheme must be chosen to handle the 
ultraviolet divergences of the theory.
In this paper we shall use dimensional regularization. The use of 
dimensional regularization in a theory that explicitly depends on 
epsilon tensors involves adopting a complicated form for the gauge 
field propagator, as will be discussed below.  In spite of this 
complication, dimensional regularization is simpler
than the Pauli-Villars regularization adopted 
in the previous paper$\cite{ref1}$ 
since it allows us to preserve explicit gauge symmetry. 

There are several problems involved with analytic continuation 
to $n$ dimensions.  The first of these is standard. The mass dimensions
of the fields and parameters become,
\begin{eqnarray}
[\phi]=[A_{\mu}]=[v]=M^{(n-2)/2};~~~[e]=M^{(3-n)/2},
~~~[\gamma_0]=M^{(2-n)/2},~~~[\lambda]=M^{2(3-n)},
\end{eqnarray}
and thus, in order to ensure that the parameters keep their 
original mass dimensions, one must make the following replacements 
for the parameters in the regularized Lagrangian,
 \begin{eqnarray}
v{\rightarrow}\mu^{(n-3)/2}v, ~~e{\rightarrow}\mu^{(3-n)/2}e, ~~
\gamma_0{\rightarrow}\mu^{(3-n)/2}\gamma_0, 
~~\lambda{\rightarrow}\mu^{2(3-n)}\lambda.
\end{eqnarray}
The second problem is more complicated. Dimensional regularization 
in a theory with a three-dimensional antisymmetric tensor
$\epsilon_{\mu\nu\rho}$ must be handled carefully. It has been 
explicitly shown that naive dimensional regularization schemes 
cannot make the theory well defined when they are applied
to a Chern-Simons type model$\cite{ref9}$. Therefore, 
in carrying out dimensional regularization we must adopt 
the three-dimensional analogue of the consistent definition for
$\gamma_5$, which was originally proposed by 't Hooft and 
Veltman$\cite{ref10}$, and later given a strict mathematical justification by
Breitenlohner and Maison$\cite{ref11}$. The explicit definition of
this dimensional continuation for Chern-Simons-type theory was
explained in Ref.$\cite{ref12}$ where it is shown that this regularization
method is indeed compatible with the Slavnov-Taylor identities.
The explicit definition for the dimensional continuation
of the epsilon and the metric tensors is
\begin{eqnarray}
\epsilon^{\mu\nu\rho} &=& 
\left\{ \begin{array}{l}
\pm 1 \,\,\,{\rm if}\,\,\, (\mu\nu\rho)={\rm permutation}\,\,\,{\rm of} \,\,\,(0,1,2) \\
0\,\,\,{\rm otherwise}
\end{array}\right. \nonumber\\
g_{\mu\nu} &=& \left\{ 
\begin{array}{l}
+1 \,\,\,{\rm for} \,\,\,\mu=\nu=0 \\
-1 \,\,\,{\rm for}\,\,\,\mu=\nu=1,2,\cdots n-1
\end{array} \right.\\
\tilde{g}_{\mu\nu} &=& \left\{
\begin{array}{l}
+1 \,\,\,{\rm for}\,\,\,\mu=\nu=0 \\  
-1 \,\,\,{\rm for} \,\,\,\mu=\nu=1,2.
\end{array}\right.  \nonumber
\end{eqnarray}
These definitions give rise to the following contractions,
\begin{eqnarray}
&&\epsilon^{\mu\nu\rho}\epsilon^{\lambda\eta}_{~~\rho}
=\tilde{g}^{\mu\lambda}\tilde{g}^{\nu\eta}
-\tilde{g}^{\mu\eta}\tilde{g}^{\nu\lambda}, 
~~~~g^{\mu\nu} \tilde{g}_\nu^{~\lambda}
=\tilde{g}^{\mu\lambda},\nonumber\\
&&\tilde{p}^\mu = \tilde{g}^{\mu\nu} p_\nu,
~~~~\tilde{p^2}=\tilde{p}_\mu p^\mu.   \nonumber
\end{eqnarray}

\subsection{Feynman rules}

The tree-level Feynman rules can be derived by standard functional integration
techniques. The propagators for the scalar fields $\chi$ and $\eta$ 
have the same form as in the four-dimensional case,
\begin{eqnarray}
iS_{\chi}(p)&=&\frac{i}{p^2-m_{\chi}^2},
 ~~~~m_{\chi}^2=\frac{\lambda}{4!}v^4;\nonumber\\ 
iS_{\eta}(p)&=&\frac{i}{p^2-m_{\eta}^2},~~~~m_{\eta}^2=\frac{\lambda}{5!}v^4.
\label{eqsp}
\end{eqnarray}

Following Ref.$\cite{ref6}$ we can obtain the dimensional regularized 
propagator for the gauge field, 
\begin{eqnarray}
&&iD_{\mu\nu}(p)=i\left\{-\left[\frac{(\gamma_0ev^2)^2}
{(p^2-e^2v^2)[(p^2-e^2v^2)^2-p^2(\gamma_0ev^2)^2]}+\frac{1}{p^2(p^2-e^2v^2)}
\right]\left(\tilde{p}^2\tilde{g}_{\mu\nu}-\tilde{p}_{\mu}
\tilde{p}_{\nu}\right)\right.\nonumber\\
&&+\frac{\gamma_0ev^2}{(p^2-e^2v^2)^2-p^2(\gamma_0ev^2)^2}
\epsilon_{\mu\nu\rho}ip^{\rho}-\frac{1}{p^2(p^2-e^2v^2)}\left[
(p^2g_{\mu\nu}-p_{\mu}p_{\nu})
-(\tilde{p}^2\tilde{g}_{\mu\nu}-\tilde{p}_{\mu}
\tilde{p}_{\nu})\right]\nonumber\\
&&+\frac{(\gamma_0ev^2)^3(p^2-\tilde{p}^2)}
{[(p^2-e^2v^2)^2-\tilde{p}^2(\gamma_0ev^2)^2]
[(p^2-e^2v^2)^2-p^2(\gamma_0ev^2)^2]}\left[\frac{\gamma_0ev^2}{p^2-e^2v^2}
(\tilde{p}^2\tilde{g}_{\mu\nu}-\tilde{p}_{\mu}
\tilde{p}_{\nu})\right.\nonumber\\
&&\left.\left.-\epsilon_{\mu\nu\rho}ip^{\rho}\right]\right\}.
\end{eqnarray}
The above expression can be rewritten as
\begin{eqnarray}
&&iD_{\mu\nu}(p)=i\left\{-\left[\frac{1}{m_1+m_2}\left(\frac{1}{m_1}
\frac{1}{p^2-m_1^2}
+\frac{1}{m_2}\frac{1}{p^2-m_2^2}\right)-\frac{1}{m_3^2}\frac{1}{p^2}\right]
\left(\tilde{p}^2\tilde{g}_{\mu\nu}-\tilde{p}_{\mu}
\tilde{p}_{\nu}\right)\right.\nonumber\\
&&+\frac{1}{m_1+m_2}\left(\frac{1}{p^2-m_2^2}-\frac{1}{p^2-m_1^2}\right)
\epsilon_{\mu\nu\rho}ip^{\rho}\nonumber\\
&&-\frac{1}{m_3^2}\left(\frac{1}{p^2-m_3^2}-\frac{1}{p^2}\right)
\left[(p^2g_{\mu\nu}-p_{\mu}p_{\nu})
-(\tilde{p}^2\tilde{g}_{\mu\nu}-\tilde{p}_{\mu}
\tilde{p}_{\nu})\right]\nonumber\\
&&\left.+\frac{(\gamma_0ev^2)^3(p^2-\tilde{p}^2)}
{[d^2-\tilde{p}^2(\gamma_0ev^2)^2][d^2-{p}^2(\gamma_0ev^2)^2]}
\left[\frac{\gamma_0ev^2}{p^2-e^2v^2}
(\tilde{p}^2\tilde{g}_{\mu\nu}-\tilde{p}_{\mu}
\tilde{p}_{\nu})-\epsilon_{\mu\nu\rho}ip^{\rho}\right]\right\}
\nonumber\\
&{\equiv}&
-i\left\{
A(p)\left(\tilde{p}^2\tilde{g}_{\mu\nu}-\tilde{p}_{\mu}
\tilde{p}_{\nu}\right)-B(p)\epsilon_{\mu\nu\rho}ip^{\rho}
+C(p)\left[(p^2g_{\mu\nu}-p_{\mu}p_{\nu})
-(\tilde{p}^2\tilde{g}_{\mu\nu}-\tilde{p}_{\mu}
\tilde{p}_{\nu})\right]\right.\nonumber\\
&&\left.+\frac{(\gamma_0ev^2)^3(p^2-\tilde{p}^2)}
{[d^2-\tilde{p}^2(\gamma_0ev^2)^2][d^2-{p}^2(\gamma_0ev^2)^2]}
\left[\frac{\gamma_0ev^2}{p^2-e^2v^2}
(\tilde{p}^2\tilde{g}_{\mu\nu}-\tilde{p}_{\mu}
\tilde{p}_{\nu})-\epsilon_{\mu\nu\rho}ip^{\rho}\right]\right\},
\label{prop2}
\end{eqnarray} 
where we have defined
\begin{eqnarray}
m_1&=&\frac{1}{2}ev\left(\sqrt{(\gamma_0v)^2+4}+\gamma_0v\right),
\nonumber\\
m_2& = &\frac{1}{2}ev\left(\sqrt{(\gamma_0v)^2+4}-\gamma_0v\right),
\label{masses}\\
m_3^2& = & e^2 v^2, ~~~~d = p^2-e^2v^2, \nonumber
\end{eqnarray}    
and 
\begin{eqnarray}
A(p)& = &\frac{1}{m_1+m_2}\left(\frac{1}{m_1}\frac{1}{p^2-m_1^2}
+\frac{1}{m_2}\frac{1}{p^2-m_2^2}\right)-\frac{1}{m_3^2}\frac{1}{p^2},
\nonumber\\
B(p)& = &\frac{1}{m_1+m_2}\left(\frac{1}{p^2-m_1^2}-\frac{1}{p^2-m_2^2}
\right),\label{ABC1}\\
C(p)& = &\frac{1}{m_3^2}\left(\frac{1}{p^2-m_3^2}
-\frac{1}{p^2}\right).\nonumber
\end{eqnarray}
To further simplify the notation we make the following definitions,
\begin{eqnarray} S(p) &=& \frac{1}{p^2}, 
~~~S_i(p) = \frac{1}{p^2 - m_i^2};\,\,\,\,\,\,i=1,2,3, \nonumber
\end{eqnarray}
which allow us to write,
\begin{eqnarray}
A(p) &=& \frac{1}{m_1m_2(m_1+m_2)}[m_2S_1 + m_1S_2 - (m_1+m_2)S], \nonumber \\
B(p) &=& \frac{1}{m_1+m_2} [S_1-S_2], 
~~~C(p) = \frac{1}{m_1m_2}(S_3-S). 
\label{ABC2}
\end{eqnarray}
These expressions obey the following identities:
\begin{eqnarray}
&&2e^2v^2(A-C) + e\gamma_0v^2B = S_1+S_2+S_3, \nonumber \\
&& e^2v^2(A-C) + e\gamma_0v^2B = \frac{1}{m_1+m_2}[m_1S_1 + m_2S_2 
- (m_1+m_2)S_3], \label{tricks}\\
&&A-C = \frac{1}{m_1m_2(m_1+m_2)}[m_2S_1 + m_1S_2 - (m_1+m_2)S_3], \nonumber \\
&& e^2A + \frac{1}{2} e\gamma_0 B = \frac{1}{2v^2}(S_1 + S_2 - 2S). \nonumber
\end{eqnarray}
These expressions are useful to simplify calculations. 

There are several mass poles in the dimensional regularized 
gauge field propagator. The
first two terms in (\ref{prop2}) show that $m_1$ and $m_2$ 
are the  photon masses
in the original three-dimensional space-time. The third term indicates 
that $m_3$ is the mass that photons acquire in  $n-3$ dimensional
space-time. The last term is proportional to an evanescent quantity,
$p^2-\tilde{p}^2$, and the power-counting shows that 
this term behaves as $1/p^5$ for large $p$. In the two-loop calculation 
the contribution of this term at the level of regularization is finite 
and hence vanishes in the limit $n\longrightarrow 3$.

The interaction vertices of 
the model were derived 
in Ref.$\cite{ref1}$. These vertices are shown in Figs. 1-3 with dotted 
lines denoting gauge bosons, while solid
lines corresponding to either the  $\eta$ field or the $\chi$ field, as labeled. 
For conciseness, the vertices in the figures are labeled purely by their number, so that for example $3b$ denotes $V_{3b}$, etc.
The Feynman rules for the vertices and the corresponding interaction
Lagrangians are as follows: 
\begin{eqnarray}
&& {\cal L}^{(1)} \longrightarrow V_{1a}(\chi^6) = -i\lambda, \,\,\,\,\,\,\,\,V_{1b}(\chi^4 \eta^2) = V_{1b}(\chi^2\eta^4) = -i\lambda/5; \nonumber \\
&&{\cal L}^{(2)} \longrightarrow V_{2a}(\chi^5) = -i\lambda v,
\,\,\,\,\,\,\,\, V_{2b}(\chi^3 \eta^2) = V_{2c}(\chi\eta^4) 
= -i\lambda v /5; \nonumber \\
&&{\cal L}^{(3)} \longrightarrow V_{3a}(\chi^4) = -i\lambda v^2/2,\,\,\,\,
\,\, \,\,\,\,\,\,\,\,V_{3b}(\chi^2\eta^2) = V_{3b}(\eta^4)
= -i\lambda v^2/10; \nonumber\\
&& {\cal L}^{(4)} \longrightarrow V_{4a}(\chi^3) = -i\lambda v^3/6,\,\,\,\,\,
\,\,\,\,\,\,\,\,\,V_{4b}(\chi\eta^2) = -i\lambda v^3 / 30;
 \label{vertices}\\
&&{\cal L}^{(5)} + {\cal L}^{(6)} \longrightarrow V_5(\eta\chi A) 
+ V_6(\eta\chi A) = e(p-q)_\lambda + 
i\gamma_0\epsilon^{\tau\alpha\lambda}k_\tau p_\alpha; \nonumber \\
&& {\cal L}^{(7)} + {\cal L}^{(8)}\longrightarrow V_7(\chi^2 A^2)
+V_8(\chi^2 A^2) =  V_7(\eta^2 A^2) + V_8(\eta^2 A^2) =
 2ie^2g_{\alpha\beta}-e\gamma_0 v
\epsilon_{\alpha\lambda\beta}(p-q)_\lambda; \nonumber \\
&&{\cal L}^{(9)} + {\cal L}^{(10)} \longrightarrow V_9(\chi A^2) 
+ V_{10}(\chi A^2)= - e\gamma_0 v\epsilon_{\mu\alpha\nu}(k-q)^\alpha 
+ 2ie^2vg_{\mu\nu}.\nonumber
\end{eqnarray}

\section{Two-loop Renormalizability}

In this section we discuss the renormalizability of the theory at the 
two-loop level. In order to show that the theory is renormalizable we 
must show that contributions to the two-loop quantum effective action 
from terms that do not appear in the original Lagrangian are finite. The superficial degree of divergence (SDD) of a diagram is given by 
\begin{eqnarray}
\omega = 3L + \sum_v \delta_v v_v -2I ,
\end{eqnarray}
where $L$ is the number of loops, $I$ is the number of internal lines, $\delta_v$ is the number of derivatives associated with each vertex, and  $v_v$ is the number of vertices of each type. The 
sum is over all vertices in the diagram.  We can rewrite 
this expression using the following relations,
\bea 
&&L=I+1-V, \nonumber \\ 
&& I=\frac{1}{2}\sum_v i_v v_v, \label{x1} \\
&& l_v = i_v + e_v, 
 \nonumber
\eea
where $V=\sum_v v_v$ is the number of vertices, $l_v$ is the total number of 
lines entering vertex $v$, and $i_v$ and $e_v$ are respectively the 
number of internal and external lines entering the vertex $v$. 
Furthermore, there exists a relation between the number of vertices and lines, 
\bea
\sum_v l_v v_v = 2I + E.
\eea
Using (\ref{x1}) with $L=2$ this constraint can be written as
\bea
\sum_v(l_v -2)v_v = 2+E \label{constraint1}
\eea
Consequently, we have the SDD
\bea
\omega =3+ \sum_v \omega_v v_v - \frac{1}{2}E,
\eea
where 
$\omega_v {\equiv}\delta_v+\frac{1}{2}l_v-3$ is the degree of divergence 
for each vertex,  
which can be read out from the interaction Lagrangian (\ref{eqla}),
\bea
&&\omega_1 = 0,~~~~\omega_2=-\frac{1}{2},~~~~~\omega_3=
-1,~~~~~\omega_4=-\frac{3}{2},~~~~~\omega_5=-\frac{1}{2},\nonumber \\
&&\omega_6=\frac{1}{2},~~~~~\omega_7=-1,~~~~~\omega_8=0,~~~~~\omega_9=
-\frac{1}{2},~~~~~\omega_{10} = -\frac{3}{2}.\nonumber
\eea
Substituting in we get,
\bea
\omega = 3-\frac{1}{2}E + \left(-\frac{1}{2} v_2 - v_3 
- \frac{3}{2}v_4 - \frac{1}{2}v_5 + \frac{1}{2}v_6 - v_7 
- \frac{1}{2}v_9 - \frac{3}{2}v_{10}\right).\label{constraint2}
\eea


To explicitly verify the renormalizability, we need to look at
the 1PI parts of the Green functions that have positive SDD but no
correspondence in the classical Lagrangian. We consider the 1PI part of the  four-point function
of gauge fields, $A^4{\equiv}\langle A_{\mu}A_{\nu}A_{\lambda}A_{\rho}\rangle$.
It is straightforward to determine which combinations of vertices satisfy 
the two conditions ({\ref{constraint1}) and (\ref{constraint2}).  
One must then select those combinations from which it is possible to 
obtain a diagram with four external photon lines.  We categorize these 
diagrams in three groups: diagrams with two internal $V_6's$, one 
internal $V_6$ and no internal $V_6$.

Note that when calculating the SDD of an $A^4$ diagram, the following points must be taken into account.  If $V_6$ is an external vertex, 
then its degree of divergence is $\omega_6=-1/2$ instead of $\omega_6=1/2$. 
This change occurs because of the fact that $V_6$ carrys one factor 
of momentum from the photon line, which becomes an external momentum 
when $V_6$ is an external vertex. Similarly, the degree of divergence 
of $V_8$ is reduced to $\omega_8=-1$ instead of $\omega_8=0$ when $V_8$ 
is an external vertex ($V_8$ is considered an external vertex if both 
photon lines are external legs). These exceptions must be considered
in looking for divergent diagrams.  
 
From (\ref{constraint2}) all diagrams with 2 $V_6's$ have SDD = 2, 1, or 0.
  The diagrams with SDD equal to 0 and no external $V_8's$ are shown in 
Figs. 4-10. It is straightforward to see that if the SDD is 2, then 
the graph has two external $V_8's$. As discussed above, these 
graphs have factors of the
 external photon momentum from the vertices $V_8$ which can be factored out,
 leaving a two-loop integral with SDD =0. The diagrams are the same as those
 shown in Fig. 4, with the external $V_7's$ replaced by $V_8's$. If the SDD 
is 1, then the diagram has one external $V_8$. These graphs also have a 
factor of external photon momentum which can be factored out to leave 
an integral of SDD=0. The diagrams are the same as those shown in shown in 
Figs. 5-7 with the external $V_7's$ replaced by $V_8's$.  

The same thing happens for diagrams with one $V_6$.  The SDD is 
either 1 or 0. The diagrams with SDD equal to 0 and no external 
$V_8's$ are shown in Figs. 11, 12. If the SDD is 1, then 
the diagram has an external $V_8$ which reduces 
the SDD to 0, with a prefactor that contains an external photon 
momentum. These diagrams are the same as those shown in Fig. 11 
with the external $V_7's$ replaced by $V_8's$. For the diagrams with 
no $V_6$, all $V_8's$ contribute one photon line which is internal 
and one which is external, and thus contribute the naive value 
$\omega_8=0$ to the counting 
 of the diagram's SDD. All divergent 
diagrams have SDD=0 and are shown in Figs. 13-15. 
 
Note that all of the diagrams with external $V_5's$ have partner diagrams 
with all possible combinations and permutations of $V_5's$ replaced 
by $V_6's$.  These partner diagrams will have the same SDD=0 as the 
original diagrams, but they will contain additional factors of external 
photon momenta, since they contain external $V_6's$.  

The above analysis shows that all the two-loop divergent 
diagrams for $A^4$ have SDD=0. For the 
purpose of studying renormalizability, we need to extract only 
the divergent parts and see whether they cancel.  
There are several simplifications that one can make when extracting the divergent part of a diagram with SDD=0.   We start by considering the scalar propagator. When a scalar
propagator (\ref{eqsp}) contains an external momentum it has the form 
\bea
S(q+p,m) =  \frac{1}{(q+p)^2-m^2}\nonumber
\eea
where $p$ is an external momentum and $q$ is an internal monentum that will be integrated over.  To study the UV behaviour of this propagator we can employ following decomposition,
\begin{eqnarray}
S(q+p,m)=\frac{1}{(q+p)^2-m^2}=\frac{1}{q^2}
-\frac{2p{\cdot}q+p^2-m^2}{q^2\left[(q+p)^2-m^2\right]}.
\end{eqnarray}
The UV degree of divergence of the second term is one less than
that of the first term and thus gives no contribution to the UV divergent part of the integral. Thus, in calculating the divergent part of the integral, we can make the replacement \bea S(q+p,m){\longrightarrow}
S(q)\,{\equiv}\,1/q^2.\eea To avoid the IR divergence induced by this procedure, it is necessary to keep the mass term at least in one propagator, i.e. replace
$S(q+p,m)$ by $S(q,m)$. We will keep all three.

Similar simplifications are possible for the gauge propagator (\ref{prop2}).  Firstly, the evanescent parts of the gauge field propagator will give vanishing contribution to 
the UV divergent part of the integral in the limit $n{\rightarrow}3$ since they have good 
UV behaviour.  Secondly, the UV degree of divergence of the parity odd term in the gauge field propagator
is one less than that of the parity even part of the propagator. Thus, there are no contributions to the UV divergent part of the integral from terms in the propagator that are proportional to the epsilon tensor. Lastly, as explained above for the scalar propagator, factors of external momentum can be dropped.   

In addition, the calculation is simplified because of the fact that there are no UV divergences
at one-loop in dimensional regularization. This fact is due to the
special analytic properties of the one-loop amplitudes in odd dimensional 
space-time$\cite{ref16a}$. As a consequence, there is no need for us to consider
the subtraction of sub-divergences.

Using the above observations, one can show that all divergent terms will be proportional to the two-loop integral 
of the form,
\bea
I = \int dk\,\int dq S(q)\,S(k)\,S(k+q,m) 
\eea
with $dk{\equiv}d^3k/(2\pi)^3$.  The divergence will appear as a pole of the form 
$1/(3-n)$.

We concentrate first on diagrams in which the prefactor does not depend 
on the external photon momentum (Figs. 4-15). These diagrams can be further
 classified into three groups by coupling constant prefactors. 
The first group of diagrams is shown in Figs. 4, 5, 8, 11, 13, 14 and is proportional to $e^2\gamma_0^2$, the second group is shown in Figs. 6, 7, 10, 12, 15 and is proportional to $e^4\gamma_0^2(\gamma_0 v)^2$.
The third group are shown in Fig. 9 and is proportional to $e^4\gamma_0^2(\gamma_0v)^4$.
Each of these sets of diagrams must be separately finite if the theory is 
to be renormalizable.  We will concentrate on the first group 
of diagrams, Figs. 4, 5, 8, 11, 13, 14. Direct calculation shows that 
the divergent pieces cancel between the diagrams in 
Figs. 4, 5, 8, 11, 13, 
while Fig. 14 is finite.  

As a further check, we calculate the $A^4$ term with four external 
momentum factors. Gauge invariance requires that this contribution 
should take the form $(F_{\mu\nu})^4$, and renormalizability requires 
that it be finite.  We obtain contributions from the diagrams in Fig. 8, 
with all of the external $V_5's$ replaced by $V_6's$. Direct calculation 
shows that the divergent pieces cancel.

Unfortunately, further checking reveals that the action is not in fact
 renormalizable.  We have calculated the $\chi^4$ term with four external momentum factors. The corresponding diagrams 
are shown in Fig. 16. This term is a contribution to the gauge invariant
 structure $(D_\mu \phi)^4$ and must be finite if the action is to be 
 renormalizable.  We find that in this case the divergent terms do not 
cancel so that the two-loop quantum effective action is not renormalizable.  

 However, as we will show in the next section, the two-loop effective
potential can be made renormalizable providing certain conditions are
satisfied by the coupling constants in the model. Therefore, 
for the lowest order in the momentum expansion of 
the quantum effective action,
the non-minimal model may yield physically relevant predictions for the
spontaneous symmetry breaking by the radiative corrections at the 
two-loop level. 

\section{The effective potential}

The effective potential is the energy density of the vacuum in which 
the expectation value of the scalar field is given by 
$\langle \phi\rangle =v$$\cite{ref7}$. 
It can be determined from the effective action 
$\Gamma[\tilde{\phi}]$ according to 
\begin{eqnarray}
\Gamma[\tilde{\phi}=v]=-(2\pi)^n\delta^{(n)}(0) V_{\rm eff}(v),
\end{eqnarray}
where $\tilde{\phi}$ is the vacuum expectation value of 
the scalar field in the presence of the external source.
Using the fact that $\Gamma[\tilde{\phi}]$ is the 
generating functional of the proper vertex,
\begin{eqnarray}
\Gamma[\tilde{\phi}]=\sum_{j=1}^{\infty}\frac{1}{j!}
\int d^n x_1d^n x_2{\cdots}d^nx_j \Gamma^{(j)}(x_1,{\cdots},x_j),\nonumber
\end{eqnarray}
one has
\begin{eqnarray}
V_{\rm eff}(v)=-\sum_{j=2}^{\infty}\frac{1}{j!}
\Gamma^{(j)}(0,0,{\cdots},0)v^j,
\label{eq13}
\end{eqnarray}
which  means that one can get the effective potential by calculating 
the 1PI vacuum diagrams.  
>From Eq.(\ref{eq13}) we can write$\cite{ref13}$
\begin{eqnarray}
V_{\rm eff}(v)=V^{{\rm tree}}(v)-\frac{i\hbar}{2}\int \frac{d^np}{(2\pi)^n}
\ln\det[iD^{-1}(p)]+i\hbar \bigg\langle\exp\left(\frac{i}{\hbar}d^nx 
{\cal L}_{\rm int}(\hbar^{\frac{1}{2}}\phi, v)\right)\bigg\rangle_{\rm 1PI},
\label{eq14}
\end{eqnarray}
where $D^{-1}(p)$ is the inverse propagator for each of the bosonic 
fields in the theory. $V^{\rm tree}$ is the tree potential and can be obtained directly from (1) as,
\bea
V^{\rm tree} = \frac{\lambda}{6!} v^6 
\eea
The second term of (\ref{eq14}) is the one-loop 
effective potential, 
 and the third term contains the higher order contributions. We shall use
this expression to get the two-loop contribution to the effective 
potential.

\subsection{ One-loop effective potential in dimensional regularization}
 
The one-loop effective potential of this model was computed 
in Ref.$\cite{ref1}$ 
in Pauli-Villars regularization. In order to discuss spontaneous symmetry breaking, we now re-calculate 
the one-loop effective potential 
 using dimensional regularization. 
According to (\ref{eq14}) we have
\begin{eqnarray}
V_{\rm eff}^{{\rm one-loop}}=\frac{\hbar}{2}\int
\frac{d^np}{i(2\pi)^n}\left\{\ln [iS_{\chi}^{-1}(p)]+\ln [iS_{\eta}^{-1}(p)]
+\ln\det[iD_{\mu\nu}^{-1}(p)]\right\}.
\label{eq30n}
\end{eqnarray}
>From the quadratic part 
${\cal L}^{(0)}$ of the Lagrangian (\ref{eq5}) 
we have the inverse scalar propagators 
\begin{eqnarray}
&&S_\chi^{(-1)}(p) = p^2-m_\chi^2, \nonumber \\
&&S_\eta^{(-1)}(p) = p^2-m_\eta^2, \nonumber 
\end{eqnarray} 
and the inverse gauge field propagator 
\begin{eqnarray}
D_{\mu\nu}^{-1}(p){\equiv}K_{\mu\nu}
=(-p^2+e^2v^2)g_{\mu\nu}+\left(1-\frac{1}{\xi}\right)p_{\mu}
p_{\nu}-\gamma_0ev^2\epsilon_{\mu\nu\rho}ip^{\rho},
\end{eqnarray}
which gives$\cite{ref6}$
\begin{eqnarray}
\det K_{\mu\nu}=\exp\mbox{Tr}\ln K_{\mu\nu}=\left(p^2-e^2v^2\right)
\left(e^2v^2-\frac{1}{\xi}p^2\right)\left[(p^2-e^2v^2)
-(\gamma_0ev^2)^2\tilde{p}^2\right].
\end{eqnarray}
Substituting above results into Eq.(\ref{eq30n}) and making use of 
the integration formulas given in Ref.$\cite{ref6}$,
\begin{eqnarray}
&&\int\frac{d^n p}{i(2\pi)^n}\ln (p^2-m^2)
=-\frac{\Gamma\left(-\frac{1}{2}n\right)}{(4\pi)^{n/2}}m^n
\stackrel{n{\rightarrow}3}{=}-\frac{1}{6\pi}m^3;\nonumber\\
&&\int\frac{d^n p}{i(2\pi)^n}\ln [(p^2-e^2v^2)^2-(\gamma_0ev^2)^2\tilde{p}^2]
\stackrel{n{\rightarrow}3}{=}-\frac{1}{6\pi}(ev)^3 (4+\gamma_0^2v^2)^{1/2}
(1+\gamma_0^2v^2),
\end{eqnarray}
we obtain the one-loop effective potential in the Landau gauge ($\xi=0$),
\begin{eqnarray}
V_{\rm eff}^{{\rm one-loop}}&=&-\frac{\hbar}{12\pi}\left[m_{\chi}^3
+m_{\eta}^3+(ev)^3(4+\gamma_0^2v^2)^{1/2}(1+\gamma_0^2v^2)\right]\nonumber\\
&=&-\frac{\hbar}{12\pi}\left\{\left[\left(\frac{1}{4!}\right)^{3/2}
+\left(\frac{1}{5!}\right)^{3/2}\right]\lambda^{3/2}v^6+
(ev)^3\left(4+\gamma_0^2v^2\right)^{1/2}\left(1+\gamma_0^2v^2\right)\right\}.
\end{eqnarray}
This result agrees with the one obtained in \cite{ref1}
\subsection{ Two-loop effective potential}

It is straightforward to see that the two-loop effective potential 
is given by diagrams that contain only vertices from the cubic and 
quartic parts of the interaction Lagrangian.  
All of these diagrams are either `theta' type diagrams, 
or `figure eight' type diagrams. 
For each diagram we give the integral expression for the amplitude.  
The integral is evaluated by rotating to Euclidean space and using 
the integrals given in the appendix. The theta diagrams contain ultraviolet
 divergent terms and the results are separated into divergent and 
finite pieces. The divergent piece is represented by 
the integral $I$ as defined in the appendix, and thus proportional to the
factor 
\begin{eqnarray}
\frac{1}{3-n} - \gamma +1 +{\rm ln} 4\pi\nonumber
\end{eqnarray}
where $\gamma$ is the Euler constant.
As a last step we use (\ref{masses}) to write the results in terms of the fundamental parameters $e$, $\gamma_0$ and $v$. 

\vspace{3mm}

\begin{flushleft}
{\bf (1).} The `figure eight' diagrams containing two scalar loops (Fig. 17)
\end{flushleft} 

\noindent These diagram are produced by the quartic interaction 
${\cal L}^{(3)}$. The amplitude can be calculated by rotating to 
Euclidean space.  We obtain,   
\begin{eqnarray}
V_{\rm eff}^{(1)}&=& -\hbar^2\lambda v^2\mu^{(3-n)}\int\frac{d^nq}{(2\pi)^n}
\frac{d^np}{(2\pi)^n}\left[\frac{3}{2}\frac{1}{
(p^2-m_{\chi}^2)(q^2-m_{\chi}^2)}\right.\nonumber\\
&&+\left.\frac{3}{10}\frac{1}{
(p^2-m_{\eta}^2)(q^2-m_{\eta}^2)}+\frac{1}{10}\frac{1}{
(p^2-m_{\chi}^2)(q^2-m_{\eta}^2)}\right]\nonumber\\
&=&\frac{\hbar^2\lambda v^2}{16\pi^2}\left(\frac{3}{2}m_{\chi}^2+
\frac{3}{10}m_{\eta}^2+\frac{1}{10}m_{\chi}m_{\eta} \right)\nonumber\\
&=&\frac{\hbar^2\lambda^2 v^6}{16\pi^2}\left(\frac{3}{2}\frac{1}{4!}+
\frac{3}{10}\frac{1}{5!}+\frac{1}{10}\frac{1}{\sqrt{5}}\frac{1}{4!}\right).
\end{eqnarray}

\vspace{3mm}

\begin{flushleft}
{\bf (2).} The `figure eight' diagram containing one scalar loop and 
one gauge field loop (Fig. 18)
\end{flushleft}
 
\noindent This diagram is produced by 
the quartic interactions 
${\cal L}^{(7)}$ and ${\cal L}^{(8)}$.  
Its  amplitude is
\begin{eqnarray}
V_{\rm eff}^{(2)}&=& i\hbar^2 \mu^{(3-n)}\int\frac{d^np}{(2\pi)^n}
\frac{d^nq}{(2\pi)^n} iD^{\mu\nu}(p)\left(2ie^2g_{\mu\nu}-2\gamma_0 e
\epsilon_{\mu\nu\rho}\tilde{p}^{\rho}\right)\left(\frac{1}{
q^2-m_{\chi}^2}+\frac{1}{q^2-m_{\eta}^2} \right)\nonumber\\
&=& 2e^2{\hbar}^2\mu^{(3-n)}\int\frac{d^np}{(2\pi)^n}
\frac{d^nq}{(2\pi)^n}\left\{-2\tilde{p}^2\left[\frac{1}{m_1+m_2}
\left(\frac{1}{m_1}\frac{1}{p^2-m_1^2}+\frac{1}{m_2}\frac{1}{p^2-m_2^2}
\right)-\frac{1}{m_3^2}\frac{1}{p^2}\right]\right.\nonumber\\
&&\left.-\frac{1}{m_3^2}\left(\frac{1}{p^2-m_3^2}
-\frac{1}{p^2}\right)\left[(n-1)p^2-2\tilde{p}^2\right]\right\}
\left(\frac{1}{q^2-m_{\chi}^2}+\frac{1}{q^2-m_{\eta}^2}\right)\nonumber\\
&-&4\gamma_0 e\hbar^2\mu^{(3-n)}\int \frac{d^np}{(2\pi)^n}
\frac{d^nq}{(2\pi)^n}\frac{\tilde{p}^2}{m_1+m_2}\left(\frac{1}{p^2-m_1^2}
-\frac{1}{p^2-m_2^2}\right)\left(\frac{1}{
q^2-m_{\chi}^2}+\frac{1}{q^2-m_{\eta}^2} \right).
\end{eqnarray}
Rotating to Euclidean space we obtain
\begin{eqnarray}
V_{\rm eff}^{(2)}&=& -2e^2\hbar^2 \mu^{(3-n)}\int\frac{d^np}{(2\pi)^n}
\frac{d^nq}{(2\pi)^n}\left[\frac{2}{m_1(m_1+m_2)}\frac{\tilde{p}^2}
{p^2+m_1^2}+\frac{2}{m_2(m_1+m_2)}\frac{\tilde{p}^2}
{p^2+m_2^2}\right.\nonumber\\
&&\left.+\frac{n-1}{m_3^2}\frac{p^2}{p^2+m_3^2}
-\frac{2}{m_3^2}\frac{\tilde{p}^2}{p^2+m_3^2}\right]
\left(\frac{1}{q^2+m_{\chi}^2}+\frac{1}{q^2+m_{\eta}^2} \right)\nonumber\\
&-&4\gamma_0 e\hbar^2\mu^{(3-n)}\int \frac{d^np}{(2\pi)^n}
\frac{d^nq}{(2\pi)^n}\frac{\tilde{p}^2}{m_1+m_2}\left(\frac{1}{p^2+m_1^2}
-\frac{1}{p^2+m_2^2}\right)\left(\frac{1}{q^2+m_{\chi}^2}
+\frac{1}{q^2+m_{\eta}^2} \right)\nonumber\\
&=&\frac{\hbar^2}{4\pi^2}\frac{m_\chi+m_\eta}{m_1+m_2}\left[
e^2(m_1^2+m_2^2)+\gamma_0 e (m_1^3-m_2^3)\right]\nonumber\\
&=&\frac{\hbar^2}{4\pi^2}\frac{1}{v^2}\frac{m_\chi+m_\eta}{m_1+m_2}
\left(m_1^4+m_2^4\right)
=\frac{\hbar^2}{4\pi^2}\frac{1}{v^2}\frac{m_\chi+m_\eta}{m_1+m_2}
\left[(m_1^2-m_2^2)^2+2 m_1^2 m_2^2\right]\nonumber\\
&=&\frac{\hbar^2}{4\pi^2}\frac{\sqrt{\lambda/4!}+\sqrt{\lambda/5!}}
{\sqrt{\gamma_0 v^2+4}}(ev)^3\left[4(\gamma_0 v)^4+(\gamma_0 v)^2+2\right].
\end{eqnarray}

\begin{flushleft}
{\bf (3).} The `theta' diagrams constructed from scalar 
propagators (Fig. 19)
\end{flushleft}

\noindent These diagrams come from 
the interaction ${\cal L}^{(4)}$ 
and contain UV divergences. Their contribution to the effective potential 
is 
\begin{eqnarray}
V_{\rm eff}^{(3)}&=& -\frac{\hbar^2}{2}\frac{(\lambda v^3)^2\mu^{(3-n)}}
{3!^3}\int\frac{d^np}{(2\pi)^n}
\frac{d^nq}{(2\pi)^n}\frac{1}{(p^2-m_{\chi}^2) (q^2-m_{\chi}^2)
\left[(p+q)^2-m_{\chi}^2\right]}\nonumber\\
&-& \frac{\hbar^2}{2}\frac{(\lambda v^3)^2\mu^{(3-n)}}
{(5!{\times}3!)^2{\times}2!}\int\frac{d^np}{(2\pi)^n}
\frac{d^nq}{(2\pi)^n}\frac{1}{(p^2-m_{\eta}^2) (q^2-m_{\eta}^2)
\left[(p+q)^2-m_{\chi}^2\right]}\nonumber\\
&=& \frac{\hbar^2}{2}(\lambda v^3)^2 
\left(\frac{1}{3!^3}+\frac{1}{(5{\times}3!)^2{\times}2!}
\right)\frac{1}{32\pi^2}
\left(\frac{1}{3-n}-\gamma+1+\ln 4\pi\right)\nonumber\\
&&-\frac{\hbar^2}{2}(\lambda v^3)^2 \frac{1}{32\pi^2}
\left[\frac{1}{3!^3}\ln\frac{9m_\chi^2}{\mu^2}
+\frac{1}{(5{\times}3!)^2{\times}2!}\ln\frac{(2m_{\eta}+m_\chi)^2}{\mu^2}
\right]\nonumber\\
&{\equiv}&V^{(3){\rm div}}+V^{(3){\rm finite}}.
\end{eqnarray}

\begin{flushleft}
{\bf (4).} The `theta' diagram composed of 
one gauge field propagator, 
one $\eta$ propagator and one $\chi$ propagator (Fig. 20).  
\end{flushleft}

\noindent This diagram  
arise from the cubic interaction ${\cal L}^{(5)}$ and ${\cal L}^{(6)}$ 
 and the UV divergence is present.  Its contribution to the 
effective potential is
\begin{eqnarray}
V_{\rm eff}^{(4)}&=& \frac{i\hbar^2}{2}\mu^{(3-n)}\int\frac{d^np}{(2\pi)^n}
\frac{d^nq}{(2\pi)^n}\left\{\left[e(p+2q)_{\mu}-i\gamma_0
\epsilon_{\mu\alpha\beta}q^{\alpha}p^{\beta}\right]\right.\nonumber\\
&\times&\left. \left[e(p+2q)_{\nu}-i\gamma_0
\epsilon_{\nu\gamma\delta}p^{\gamma}q^{\delta}\right]iD^{\mu\nu}(p)
iS_{\eta}(p+q)iS_{\chi}(q)\right\}.
\end{eqnarray}
Inserting  the gauge field propagator (\ref{prop2}),
employing the identities (\ref{tricks}) to simplify the expression 
and performing the following contractions,
\begin{eqnarray}
(p+2q)_{\mu}(p+2q)_{\nu}\left(\tilde{p}^2g_{\mu\nu}-\tilde{p}_{\mu}
\tilde{p}_{\nu}\right)
=4\left[\tilde{p}^2\tilde{q}^2-4\left(\tilde{p}\cdot\tilde{q}
\right)\right],\nonumber\\
(p+2q)_{\mu}(p+2q)_{\nu}\left(p^2g^{\mu\nu}-p^{\mu}p^{\nu}\right)
=4\left[p^2q^2-(p{\cdot}q)^2\right],\nonumber\\
\left[(p+2q)_{\mu}\epsilon_{\nu\gamma\delta}p^{\gamma}q^{\delta}
+\epsilon_{\mu\alpha\beta}q^{\alpha}p^{\beta}(p+2q)_{\nu}\right]
\left(\tilde{p}^2g_{\mu\nu}-\tilde{p}_{\mu}
\tilde{p}_{\nu}\right)\nonumber\\
=\left[(p+2q)_{\mu}\epsilon_{\nu\gamma\delta}p^{\gamma}q^{\delta}
+\epsilon_{\mu\alpha\beta}q^{\alpha}p^{\beta}(p+2q)_{\nu}\right]
\left(p^2g^{\mu\nu}-p^{\mu}p^{\nu}\right)=0,
\nonumber\\
\left[(p+2q)_{\mu}\epsilon_{\nu\gamma\delta}p^{\gamma}q^{\delta}
+\epsilon_{\mu\alpha\beta}q^{\alpha}p^{\beta}(p+2q)_{\nu}\right]
\epsilon^{\mu\nu\rho}p_{\rho}
=4\left[\tilde{p}^2\tilde{q}^2-4\left(\tilde{p}\cdot\tilde{q}
\right)\right]\nonumber\\
\epsilon_{\mu\alpha\beta}q^{\alpha}p^{\beta}\epsilon_{\nu\gamma\delta}
p^{\gamma}q^{\delta}\left(\tilde{p}^2g_{\mu\nu}-\tilde{p}_{\mu}
\tilde{p}_{\nu}\right)=\tilde{p}^2\left[(p{\cdot}q)^2
-\tilde{p}^2\tilde{q}^2\right], \nonumber\\
\epsilon_{\mu\alpha\beta}q^{\alpha}p^{\beta}\epsilon_{\nu\gamma\delta}
p^{\gamma}q^{\delta}\left(p^2g^{\mu\nu}-p^{\mu}p^{\nu}\right)=
p^2\left[\tilde{p}^2\tilde{q}^2-4\left(\tilde{p}\cdot\tilde{q}
\right)\right],\nonumber\\
\epsilon_{\mu\alpha\beta}q^{\alpha}p^{\beta}\epsilon_{\nu\gamma\delta}
p^{\gamma}q^{\delta}\epsilon^{\mu\nu\rho}p_\rho=0,
\end{eqnarray}
we obtain
\begin{eqnarray}
V_{\rm eff}^{(4)}&=& -\frac{\hbar^2}{2}\int\frac{d^np}{(2\pi)^n}
\frac{d^nq}{(2\pi)^n}\mu^{(3-n)}\left\{
\left[-\frac{1}{m_1 (m_1+m_2)}S_1(p)-\frac{1}{m_2 (m_1+m_2)}S_2(p)+
\frac{1}{m_3^2}S(p)\right]\right.\nonumber\\
&&{\times}\left.S_{\chi}(q)S_{\eta}(p+q)
\left(4e^2+\gamma_0^2\tilde{p}^2\right)
 \left[\tilde{p}^2\tilde{q}^2-\left(\tilde{p}\cdot\tilde{q}
\right)^2\right]\right\}\nonumber\\
&+&\frac{\hbar^2}{2}\frac{4e\gamma_0}{m_1+m_2}\int \frac{d^np}{(2\pi)^n}
\frac{d^nq}{(2\pi)^n}\mu^{(3-n)}\left[S_1(p)
-S_2(p)\right]S_{\chi}(q)S_{\eta}(p+q)
\left[\tilde{p}^2\tilde{q}^2-\left(\tilde{p}\cdot\tilde{q}
\right)^2\right]\nonumber\\
&-&\frac{\hbar^2}{2}\frac{1}{m_3^2}\int\frac{d^np}{(2\pi)^n}
\frac{d^nq}{(2\pi)^n}\mu^{(3-n)}\left[S_3(p)
-S(p)\right]S_{\chi}(q)S_{\eta}(p+q)\left\{
4e^2\left[p^2q^2-(p\cdot q)^2\right]\right.\nonumber\\
&&\left.-\left[4 e^2+\gamma_0^2 
(p^2-\tilde{p}^2)\right]\left[\tilde{p}^2\tilde{q}^2
-\left(\tilde{p}\cdot\tilde{q}\right)^2\right]\right\},
\end{eqnarray}
where 
the evanescent terms are thrown away since their contribution vanishes 
in the limit $n\rightarrow 3$. Using the integrals given in the appendix , 
we find for the divergent part, 
\begin{eqnarray}
V_{\rm eff}^{(4)({\rm div})} &=&{\hbar^2\over 64\pi^2}\left({1\over 3-n} -\gamma
+1 +\ln(4\pi)\right)\left(\left({4\over v^2}+\gamma_0^2\right)
\left({1\over2} \left((m_1-m_2)^2 + m_1m_2\right)(m_\chi^2+m_\eta^2) 
\right.\right.
\none
& &-{1\over 4}(m_1^2-m_2^2)^2+\left.\left.{1\over 4}m_1m_2(m_1-m_2)^2 - {1\over 4}m_1^2m_2^2\right)
 +{1\over 4} \gamma_0^2 (m_\chi^2-m_\eta^2)^2\right)
\none
&=& -{\hbar^2\over 230400 \pi^2}\left({1\over 3-n} -\gamma +1 +
  \ln(4\pi)\right)
   v^2\left(-90e^2\gamma_0^4v^6\lambda-450e^2\lambda v^4\gamma_0^2 
   \right.\none
& &\left.+
  900e^4\gamma_0^6v^6+6300e^4v^4\gamma_0^4+11700e^4\gamma_0^2v^2-360e^2\lambda 
v^2 +3600 e^4 +\gamma_0^2\lambda^2v^6\right) ,\label{v8}
\end{eqnarray}
and for the finite part,
\begin{eqnarray}
V_{\rm eff}^{(4)({\rm finite})}&=&
-{\hbar^2\over 64 \pi^2 v^2} (m^2_\chi-m^2_\eta)^2
\ln\left({(m_\chi+m_\eta)^2\over \mu^2}\right)\none
& &+{\hbar^2(\gamma_0^2v^2+4)m_2\over 256
  \pi^2v^2(m_1+m_2)}((m_\chi+m_\eta)^2-m_2^2)((m_\chi-m_\eta)^2-m_2^2)\ln\left({(m_\chi+m_2+m_\eta)^2\over \mu^2}\right)\none
& &+{\hbar^2(\gamma_0^2v^2+4)m_1\over 256
  \pi^2v^2(m_1+m_2)}((m_\chi+m_\eta)^2-m_1^2)((m_\chi-m_\eta)^2-m_1^2)\ln\left({(m_\chi+m_1+m_\eta)^2\over \mu^2}\right)\none
& & -{\hbar^2 (\gamma_0^2v^2+4)\over 1536 \pi^2 v^2(m_1+m_2)}\left(
 12(m_1^2+m_2^2)(m_\chi^3-m_\chi^2m_\eta-m_\eta^2m_\chi+m_\eta^3) 
+5m_1^5+5m_2^5
\right.\none
& &-(m_1^3+m_2^3)(10m_\chi^2+12m_\chi
 m_\eta + 10 m_\eta^2)+m_3^2(m_1+m_2)(10m_\chi^2+10m_\eta^2-5m_3^2)
\none
& & \left. +12(m_1^4+m_2^4)(m_\chi+m_\eta)\right)\none
& &-{\hbar^2\gamma_0^2\over 7680 \pi^2}
\left( 6m_1^4+6m_2^4 - (12
m_\eta^2+12m_\chi^2+6m_3^2)(m_1^2+m_2^2)  + 25 m_3^4\right.
\none
& &\left. -26(m_\chi^2+m_\eta^2)m_3^2 + 60m_\eta
  m_\chi(m_\chi^2+m_\eta^2) + 25(m_\chi^2-m_\eta^2)^2\right).
\end{eqnarray}

\begin{flushleft}
{\bf (5).} The `theta' diagram with one $\chi$ propagator and two gauge 
field propagators (Fig. 21).
\end{flushleft}

\noindent The interaction Lagrangians yielding this diagram 
are  ${\cal L}^{(9)}$ and ${\cal L}^{(10)}$. 
The corresponding contribution to the effective potential is, 
\begin{eqnarray}
V_{\rm eff}^{(5)}&=& i{\hbar^2}\int\frac{d^np}{(2\pi)^n}
\frac{d^nq}{(2\pi)^n}\mu^{(3-n)}\left\{\left[2ie^2v g_{\mu\lambda}
+\gamma_0 v e\epsilon_{\mu\lambda\alpha}\left(\tilde{p}^{\alpha}
-\tilde{q}^{\alpha}\right)\right]iD_{\lambda\rho}(q)\right.\nonumber\\
&&{\times}\left.iS_{\chi}(p+q)\left[2ie^2v g_{\rho\nu}
+\gamma_0 v e\epsilon_{\rho\nu\beta}\left(\tilde{p}^{\beta}
-\tilde{q}^{\beta}\right)\right] iD_{\nu\mu}(p)\right\}.
\end{eqnarray}
Denoting 
\begin{eqnarray}
V_{\mu\nu}(p,q)&{\equiv}&2ie^2v g_{\mu\nu}
+\gamma_0 v e\epsilon_{\mu\nu\rho}\left(\tilde{p}^{\rho}
-\tilde{q}^{\rho}\right), \nonumber\\
P^{(T)}_{\mu\nu}&{\equiv}&p^2g_{\mu\nu}-p_{\mu}p_{\nu},\nonumber\\
\tilde{P}^{(T)}_{\mu\nu}&{\equiv}&\tilde{p}^2
\tilde{g}_{\mu\nu}-\tilde{p}_{\mu}\tilde{p}_{\nu},
\end{eqnarray}
and performing the following contractions,
\begin{eqnarray}
V_{\mu\lambda}(p,q)V_{\rho\nu}(p,q)\tilde{P}^{(T)\mu\nu}
\tilde{P}^{(T)\lambda\rho}=&-&4e^4v^2\left[
\tilde{p}^2\tilde{q}^2+(\tilde{p}{\cdot}\tilde{q})^2\right]
+(\gamma_0 ve)^2\nonumber\\
&{\times}&\left[-(\tilde{p}^2+\tilde{q}^2)\left(
\tilde{p}^2\tilde{q}^2+(\tilde{p}\cdot\tilde{q})^2\right)
+4\tilde{p}^2
\tilde{q}^2(\tilde{p}{\cdot}\tilde{q})\right],
\nonumber\\
V_{\mu\lambda}(p,q)V_{\rho\nu}(p,q)\tilde{P}^{(T)\mu\nu}P^{(T)\lambda\rho}
&=&-4e^4v^2\left[2\tilde{p}^2q^2-\tilde{p}^2\tilde{q}^2+
\left(\tilde{p}{\cdot}\tilde{q}\right)^2\right]
+(\gamma_0 v e)^2\left[4\tilde{p}^2q^2(\tilde{p}\cdot\tilde{q})
\right.\nonumber\\
&&\left.-2\tilde{p}^4q^2+\tilde{p}^4\tilde{q}^2-\tilde{p}^2
(\tilde{p}\cdot\tilde{q})^2
-q^2(\tilde{p}\cdot\tilde{q})^2-\tilde{p}^2
\tilde{q}^2q^2\right], \nonumber\\
V_{\mu\lambda}(p,q)V_{\rho\nu}(p,q)P^{(T)\mu\nu}
P^{(T)\lambda\rho}&=&-4e^4v^2\left[(n-3)p^2q^2+p^2q^2+(p\cdot q)^2\right]
\nonumber\\
&&+(\gamma_0ve)^2\left[-2p^2q^2(\tilde{p}^2+\tilde{q}^2)+
4p^2q^2(\tilde{p}{\cdot}\tilde{q})\right.\nonumber\\
&&\left.-(p^2+q^2)(\tilde{p}
\cdot \tilde{q})^2+(p^2+q^2)\tilde{p}^2\tilde{q}^2\right],
\nonumber\\
 V_{\mu\lambda}(p,q)V_{\rho\nu}(p,q)\tilde{P}^{(T)\mu\nu}
\epsilon^{\lambda\rho\gamma}iq_{\gamma}&=&-2\gamma_0v^2e^3\left[-4
\tilde{p}^2\tilde{p}{\cdot}\tilde{q}
+2\left(\tilde{p}{\cdot}\tilde{q}\right)^2
+2 \tilde{p}^2\tilde{q}^2\right],\nonumber\\
V_{\mu\lambda}(p,q)V_{\rho\nu}(p,q)P^{(T)\mu\nu}
\epsilon^{\lambda\rho\gamma}iq_{\gamma}&=&-4\gamma_0v^2e^3\left[-2p^2
\tilde{p}{\cdot}\tilde{q}+2 p^2\tilde{q}^2+(\tilde{p}\cdot
\tilde{q})^2-\tilde{p}^2\tilde{q}^2\right],\nonumber\\
V_{\mu\lambda}(p,q)V_{\rho\nu}(p,q)\epsilon^{\mu\nu\alpha}ip_{\alpha}
\epsilon^{\lambda\rho\beta}iq_{\beta}&=&8e^4v^2\tilde{p}{\cdot}
\tilde{q}+2(\gamma_0 ve)^2\left[\left(\tilde{p}^2+\tilde{q}^2
\right)\tilde{p}\cdot\tilde{q}\right.\nonumber\\
&& \left.-\tilde{p}^2\tilde{q}^2-
(\tilde{p}\cdot\tilde{q})^2\right],
\end{eqnarray}
we get
\begin{eqnarray}
V_{\rm eff}^{(5)}&=& i{\hbar^2}\int\frac{d^np}{(2\pi)^n}
\frac{d^nq}{(2\pi)^n}\mu^{3-n}e^2v^2S_{\chi}(p+q)
\left\{e^2\left[-4e^2C(p)C(q)p^2q^2 (n-3)\right.\right.\nonumber\\
&&-16 [A(p)-C(p)]C(q)
\tilde{p}^2(q^2-\tilde{q}^2)-4C(p)C(q)[p^2q^2
+(p{\cdot}q)^2]\nonumber\\
&&\left.-4[A(p)A(q)-C(p)C(q)]
[\tilde{p}^2\tilde{p}^2+(\tilde{p}\cdot\tilde{q})^2]
+8 B(p) B(q)\tilde{p}{\cdot}\tilde{q}\right]\nonumber\\
&+&e\gamma_0\left[8A(p)B(q)[\tilde{p}^2\tilde{q}^2
+(\tilde{p}{\cdot}\tilde{q})^2-2\tilde{p}^2\tilde{p}{\cdot}
\tilde{q}]+16C(p) B(q)(p^2-\tilde{p}^2)(\tilde{q}^2-
\tilde{p}\cdot\tilde{q})\right]\nonumber\\
&+&\gamma_0^2\left[-4C(p) C(q) (p^2-\tilde{p}^2)
(q^2-\tilde{q}^2)(\tilde{p}^2-\tilde{p}{\cdot}\tilde{q})
-4A(p) C(q) (q^2-\tilde{q}^2)\tilde{p}^4\right.\nonumber\\
&&\left.- C(p) A(q)(p^2-\tilde{p}^2) (2\tilde{p}^2\tilde{q}^2
+2(\tilde{p}{\cdot}\tilde{q})^2-8\tilde{q}^2
\tilde{p}{\cdot}\tilde{q}\right.\nonumber\\
&&+2B(p) B(q)
[(\tilde{p}^2+\tilde{q}^2)\tilde{p}{\cdot}\tilde{q}-
\tilde{p}^2\tilde{q}^2-(\tilde{p}{\cdot}\tilde{q})^2]
\nonumber\\
&&\left.\left.
-2A(p)A(q)[2\tilde{p}^2\tilde{q}^2
\left(\tilde{p}^2-(\tilde{p}{\cdot}\tilde{q})\right)
-\tilde{p}^2(\tilde{p}^2\tilde{q}^2
-\left(\tilde{p}\cdot\tilde{q})^2\right)]\right]\right\}.
\label{eq34}
\end{eqnarray}
In deriving (\ref{eq34}) we have used the fact that the integrand 
is invariant 
under the interchange $p{\longleftrightarrow}q$. 

The calculation of this amplitude is quite 
lengthy. 
For clarity we divide the result in three parts.  We label them $V_{e^2}$, $V_{e\gamma_0}$ 
and $V_{\gamma_0^2}$. They correspond to the terms proportional to $e^2$,
$e\gamma_0$ and $\gamma_0^2$ respectively in (\ref{eq34}) above. 
In the expressions that appear below, the couplings $e$ and $\gamma_0$ do not 
explicitly appear because we have used the relations (\ref{masses}) to write
\begin{eqnarray}
&&e^2 = \frac{1}{v^2}m_3^2, \nonumber \\
&& e\gamma_0 = \frac{1}{v^2} (m_1-m_2), \\
&&\gamma_0^2 = \frac{1}{v^2}\frac{(m_1-m_2)^2}{m_1m_2}, \nonumber
\end{eqnarray}
and eliminate the couplings in favour of the masses. This substitution 
allows us to combine all three terms to obtain the simplest expression 
possible for the final result of this diagram. As a last step we use
 (\ref{masses}) to write the result in terms of the fundamental 
parameters $e$, $\gamma_0$ and $v$. First, the $e^2$ part is 
\begin{eqnarray}
V_{e^2}&=&-{3\hbar^2\mu^{(2n-6)}\over 16\pi^2 v^2}\left({1\over 3-n} -\gamma
+1 +\ln(4\pi)\right)  m_3^4 +{\hbar^2 m_\chi^4\over32 \pi^2 v_2} \ln\left({m_\chi^2\over
     \mu^2}\right) \none
& &
  -{\hbar^2m_1(m_2^2-m_\chi^2)^2\over 16\pi^2 (m_1+m_2) v^2}
   \ln\left({(m_2+m_\chi)^2\over \mu^2}\right) -{\hbar^2 m_2(m_1^2-m_\chi^2)^2\over 16 \pi^2(m_1+m_2)v^2}
    \ln\left({(m_1+m_\chi)^2\over \mu^2}\right)
\none
 & & +{\hbar^2m_1^2(4m_2^2-m_\chi^2)^2\over 32 (m_1+m_2)^2 \pi^2
   v^2}\ln\left( {(2m_2+m_\chi)^2\over \mu^2}\right) +{\hbar^2m_2^2(4m_1^2-m_\chi^2)^2\over 32 (m_1+m_2)^2 \pi^2
   v^2}\ln\left( {(2m_1+m_\chi)^2\over \mu^2}\right)\none
& &+{\hbar^2 m_3^2 ((m_1-m_2)^2-m_\chi^2)^2\over 16(m_1+m_2)^2\pi^2 v^2}
   \ln\left({(m_1+m_2+m_\chi)^2\over \mu^2}\right)\none
& & +{\hbar^2 m_3^4(-3(m_1-m_2)^2 + 2m_\chi(m_1+m_2+m_\chi))\over
   8 (m_1+m_2)^2 \pi^2 v^2}\label{I};
\end{eqnarray} 
The term proportional to $e\gamma_0$ is given by
\begin{eqnarray}
V_{e\gamma_0}&=& -{5\hbar^2\mu^{(2n-6)}\over 8\pi^2 v^2}  \left({1\over 3-n} -\gamma
+1 +\ln(4\pi)\right) (m_1-m_2)^2m_3^2\none
& &-{\hbar^2 m_3^2 (m_1-m_2)^2 (49m_1^2 + 49m_2^2 - 42(m_1+m_2) m_\chi +
  26 m_3^2 - 12 m_\chi^2) 
\over 48 (m_1+m_2)^2 \pi^2 v^2}\none
& & -{ \hbar^2(4 m_1^2-m_\chi^2)^2 (m_2-m_1)m_2\over 16(m_1+m_2)^2\pi^2
  v^2}
   \ln\left({(2m_1+m_\chi)^2\over \mu^2}\right)\none
& & + { \hbar^2(4 m_2^2-m_\chi^2)^2 (m_2-m_1)m_1\over 16(m_1+m_2)^2\pi^2
  v^2}
   \ln\left({(2m_2+m_\chi)^2\over \mu^2}\right)\none
& & +{\hbar^2(m_1-m_2)^2((m_1-m_2)^2-m_\chi^2)^2\over
    16(m_1+m_2)^2\pi^2 v^2}
    \ln\left({(m_1+m_2+m_\chi)^2\over \mu^2}\right)\none
& & +{\hbar^2(m_1-m_2)(m_2^2-m_\chi^2)^2 \over 16\pi^2(m_1+m_2)v^2}
  \ln\left({(m_2+m_\chi)^2\over \mu^2}\right)\none
& & -{\hbar^2(m_1-m_2)(m_1^2-m_\chi^2)^2 \over 16\pi^2(m_1+m_2)v^2}
  \ln\left({(m_1+m_\chi)^2\over \mu^2}\right)\label{II};
\end{eqnarray} 
The term proportional to $\gamma_0^2$ can be calculated in a similar
way, 
\begin{eqnarray}
V_{\gamma_0^2}&=& -{\hbar^2(m_1-m_2)^2\over 64 \pi^2 v^2}\left[ \mu^{(2n-6)}  \left({1\over 3-n} -\gamma
+1 +\ln(4\pi)\right)(25(m_1^2+m_2^2)-35m_3^2-6m_\chi^2)\right.
\none
& &-{2(4m_1^2-m_\chi^2)^2\over (m_1+m_2)^2}
    \ln\left({(2m_1+m_\chi)^2\over \mu^2}\right)-{2(4m_2^2-m_\chi^2)^2\over (m_1+m_2)^2}
    \ln\left({(2m_2+m_\chi)^2\over \mu^2}\right)\none
& &+{(m_1^2-m_\chi^2)^2 m_1\over (m_1+m_2)m_3^2}
\ln\left({(m_1+m_\chi)^2\over \mu^2}\right)+{(m_2^2-m_\chi^2)^2 m_2\over (m_1+m_2)m_3^2}
\ln\left({(m_2+m_\chi)^2\over \mu^2}\right)\none
& &-\left.{(m_1-m_2)^2((m_1-m_2)^2-m_\chi^2)^2\over (m_1+m_2)^2m_1m_2}
\ln\left({(m_1+m_2+m_\chi)^2\over \mu^2}\right)\right.\none
& & +{1\over 2310 (m_1+m_2)^2}
  \left( 89981(m_1^4+m_2^4) - 110880m_\chi(m_1^3+m_2^3)
\right.\none
& & -(3837m_3^2+28158m_\chi^2)(m_1^2+m_2^2)
+(-46200m_\chi m_3^2+9240m_\chi^3)(m_1+m_2)\none
& & \left.\left.-19356m_\chi^2m_3^2
 +34124m_3^4\right)\right].
\label{III}
\end{eqnarray}

The final expression for $V_{\rm eff}^{(5)}$ is obtained by combining~(\ref{I}),~(\ref{II}), and~(\ref{III}):
\begin{equation}
V_{\rm eff}^{(5)} = V_{e^2}+V_{e\gamma_0}+V_{\gamma_0^2}
\label{V5 total}
\end{equation}

To summarize, in the $\overline{MS}$ scheme, the effective potential to two loops is given by,
\begin{eqnarray}
V_{\rm eff}&=&V^{\rm tree}+V^{\rm one-loop}_{\rm eff}+
V^{\rm two-loop}_{\rm eff},\nonumber\\
V^{\rm two-loop}_{\rm eff}&{\equiv}&V_{\rm eff}^{(1)}+V_{\rm eff}^{(2)}+
V_{\rm eff}^{(3){\rm finite}}+V_{\rm eff}^{(4){\rm finite}}
+V_{\rm eff}^{(5){\rm finite}}.
\end{eqnarray} 
Having obtained an expression for the effective potential, the next step 
is to absorb the divergent terms into renormalized coupling constants
by choosing certain renormalization conditions. 
However, from (\ref{v8}), it is easy to see that 
$V_{\rm eff}^{(4){\rm div}}$ contains terms proportional to $v^8$. 
Since these terms would require counterterms that have
 no correspondence in the classical Lagrangian, 
we must require that they vanish to ensure the renormalizability
of two-loop effective potential. 
This can be done by imposing the constraint,
\begin{equation}
\gamma_0^2(-90e^2\gamma_0^2\lambda+900e^4\gamma_0^4+\lambda^2)=0
\end{equation}  
The solutions to this equation are
\begin{eqnarray}
\gamma_0^2=0 ~~~\mbox{or}~~~\gamma_0^2={(3\pm\sqrt{5})\lambda\over 60 e^2}.
\label{coupling condition}
\end{eqnarray}
The first, trivial solution, merely verifies that the theory is 
renormalizable in the limit $\gamma_0\rightarrow 0$. For the case of 
a finite non-minimal Chern-Simons coupling, we must choose 
one of the  second solutions and impose a non-trivial relation between this 
non-minimal coupling and the other fundamental coupling 
constants of the theory. Note that our result 
that the non-minimal coupling depends on the fundamental couplings 
is consistent with the fact that in a theory without non-minimal 
Chern-Simons coupling
the magnetic moment interaction would be generated through quantum corrections$\cite{ref14}$. We should emphasize that Eq.[\ref{coupling condition}]
applies to the tree level couplings. Since it is not protected by any symmetry, it will probably be subject to quantum corrections. However, since we
cannot go beyond two loops in this model, these corrections, which would affect the effective potential only at three loops,
 are not relevant.

\section{Summary and Conclusions}

We have studied in some details the two-loop quantum behaviour 
of $2+1$-dimensional scalar QED with a non-minimal Chern-Simons coupling. 
As expected from dimensional arguments, the complete theory 
is not renormalizable beyond the one-loop 
level. This is not a major problem in principle, since the theory is 
most reasonably considered as an effective field theory describing magnetic
 moment interactions between the charged scalar and the electromagnetic fields.
At this level the renormalizability is useful because it limits the number 
of parameters that must be introduced at each order in 
perturbation theory in order to make unambiguous predictions. 
We find  
that the effective potential, 
which is the zeroth order contribution to the quantum effective action 
in the momentum expansion, can be made renormalizable providing the 
fundamental couplings are related to the non-minimal one 
by the condition (\ref{coupling condition}). 
It is of interest to further examine the effective potential with this condition imposed. The renormalization of the two-loop effective 
potential using a physical renormalization scheme is a straightforward 
but lengthy procedure. This, plus an analysis of the resulting
symmetry breaking mechanism, will be the subject of a future publication.

Finally, we remark that the two-loop calculations presented above are
extremely lengthy, in part due to the complicated form of the dimensionally
regulated gauge field propagator. It would therefore be of interest to examine
an alternative regularization scheme that preserves gauge invariance without
the need to dimensionally continue the antisymmetric tensor. One excellent
candidate is operator regularization, which was shown to greatly facilitate 
two-loop calculations in Chern-Simons theory\cite{mckeon}.

\acknowledgments
 This work is supported by the Natural Sciences and Engineering Research
Council of Canada. We are greatly indebted  to  
 R. Kobes for various useful discussions.

\appendix

\section{Two-loop integration formulae}

This appendix is a collection of the integration formulae used in
the calculation of two-loop effective potential.
Denoting $dk = {d^nk}/{(2\pi)^n} $ and 
 using the following notation for the Euclidean space propagators:  
\begin{eqnarray}
{\cal S}_{1}(q) = \frac{1}{q^2 + m_1^2},  
~~~{\cal S}_{2}(k) &=& \frac{1}{k^2 + m_2^2},  
~~~{\cal S}_3(k+q) = \frac{1}{(k+q)^2 + m_3^2}, \nonumber\\ 
S^3 &{\equiv}& {\cal S}_{1}(q) {\cal S}_{2}(k) {\cal S}_3(k+q),  \nonumber
\end{eqnarray}
we introduce the following integrals:
\begin{eqnarray}
&&I=\int dk\,dq\, S^3; \nonumber\\
&&N_1 = \int dk\,{\cal S}(k), 
~~~N_2 = \int dk\, k^2 {\cal S}(k), 
~~~N_3 = \int dk\, k^4 {\cal S}(k); \nonumber \\
&&K^{\mu\nu\lambda\tau} = \int dk \,dq\,q^\mu q^\nu  k^\lambda k^\tau S^3;
~~~T^{\mu\nu\lambda\tau} = \int dk \,dq\,q^\mu q^\nu  q^\lambda k^\tau S^3; \nonumber \\
&&L^{\mu\nu\alpha\beta\lambda\tau} = \int dk \,dq\,q^\mu q^\nu q^\alpha 
q^\beta k^\lambda k^\tau S^3; 
~~~R^{\mu\nu\alpha\beta\lambda\tau} = \int dk \,dq\,q^\mu q^\nu 
q^\alpha k^\beta k^\lambda k^\tau S^3. \nonumber 
\end{eqnarray}
Contracting the tensor indices in various combinations with
the $n$- and three-dimensional metric tensors, we get the integrations
needed for two-loop calculations, 
\begin{eqnarray}
&&K_1 = \int dk\,dq\,q^2 k^2 S^3\,,~~~~~~~~
K_2 = \int dk\,dq\,(k\cdot q)^2 S^3\, ,~~~~~~~~{K'}_1=\int dk \,dq \,\tilde q^2 k^2 S^3\nonumber \\
&& \tilde{K}_1 = \int dk\,dq\,\tilde{q}^2\tilde{k}^2 S^3\, ,~~~~~~~~
 \tilde{K}_2 = \int dk\,dq\,(\tilde{k}\cdot \tilde{q})^2 S^3\,;\nonumber \\
&&T_1 = \int dk \,dq\,q^2 (k\cdot q) S^3\, ,~~~~~~
\tilde{T}_1 = \int dk \,dq\,\tilde{q}^2 (\tilde{k}\cdot \tilde{q}) 
S^3\, , ~~~~~~~
T_2 = \int dk \,dq\,q^2 (\tilde{k}\cdot \tilde{q}) S^3\,; \nonumber \\
&& L_1 = \int dk \,dq\,q^4k^2 S^3\, ,~~~~~~~
\tilde{L}_1 = \int dk \,dq\,\tilde{q}^4 \tilde{k}^2 S^3 \, ,~~~~~~~
L_2 = \int dk \,dq\,q^2 (k\cdot q)^2 S^3 \, ,\nonumber \\
&& \tilde{L}_2= \int dk \,dq\,\tilde{q}^2 (\tilde{k}\cdot\tilde{q})^2 S^3 
\, ,~~~~~~~ L_3 = \int dk \,dq\, q^2 \tilde{q}^2 \tilde{k}^2 S^3 \, ,~~~~~~~ 
L_4 = \int dk \,dq\, \tilde{q}^4 k^2 S^3 \,,\nonumber \\
&& L_5 = \int dk \,dq\, q^2 \tilde{q}^2 k^2 S^3 \, ,~~~~~~~ 
L_6 = \int dk \,dq\, q^2 (\tilde{k}\cdot \tilde{q})^2 S^3 \,;\nonumber\\
&& R_1 = \int dk \,dq\,q^2 k^2 (q\cdot k) S^3 \,,~~~~~~~
 R_2 = \int dk \,dq\, (q\cdot k)^3 S^3 \,,~~~~~~~
R_3 = \int dk \,dq\, \tilde{q}^2 \tilde{k}^2 (\tilde{k} \cdot \tilde{q})S^3
\nonumber\\
&& R_4 = \int dk \,dq\, q^2 \tilde{k}^2 (\tilde{k} \cdot \tilde{q}) S^3 \, ,
~~~~~~
R_5 = \int dk \,dq\, q^2 k^2 (\tilde{k} \cdot \tilde{q}) S^3 \,.\nonumber 
\end{eqnarray}
The results are listed below,
\begin{eqnarray}
&&I=\frac{\mu^{2(n-3)}}{32\pi^2} \left[ \frac{1}{3-n} -\gamma +1 
- {\rm ln}\frac{(m_1+m_2+m_3)^2}{4\pi\mu^2}\right];\nonumber \\
&&N_1 = -\frac{\mu^{2(n-3)}}{4\pi}m, 
~~~~N_2 = \frac{\mu^{2(n-3)}}{4\pi}m^3, 
~~~~N_3 = -\frac{\mu^{2(n-3)}}{4\pi}m^5; \nonumber \\
&&K_1 = -\frac{\mu^{2(n-3)}}{16\pi^2}(m_1^3 + m_2^3) m_3 
+ m_1^2m_2^2 I, \nonumber \\
&&K_2 = \frac{\mu^{2(n-3)}}{64\pi^2}\left[m_3^2\left(m_1m_3 + m_2m_3 
- m_1 m_2\right)- m_3\left(3m_1^3 + 3m_2^3 + m_1^2 m_2 
+ m_2^2 m_1 \right) \right.\nonumber \\
&&~~~~ \left.+ (m_1^2 + m_2^2)m_1m_2\right] 
+\frac{1}{4}(m_1^2 + m_2^2 - m_3^2)^2 I, \nonumber \\
&&{K'}_1 = \frac{3}{n} K_1 \nonumber \\
&&\tilde{K}_1 = \frac{1}{n(n-1)(n+2)}[(9n+3)K_1 + (6n-18)K_2], 
\nonumber \\
&&\tilde{K}_2 = \frac{1}{n(n-1)(n+2)}[3(n-3)K_1 + 6(2n-1)K_2]; \nonumber \\
&& T_1 = \frac{\mu^{2(n-3)}}{32\pi^2}[m_1^3(m_3-m_2)+ 
2m_2^3m_3 + m_1^2 m_2 m_3] - \frac{1}{2} m_1^2 
(m_1^2 + m_2^2 - m_3^2) I, \nonumber\\
&&\tilde{T}_1 = \frac{15}{n(n+2)}T_1, 
~~~~T_2 = \frac{3}{n} T_1; \nonumber \\
&&L_1 =  \frac{1}{16\pi^2}\mu^{2(n-3)} \left[ m_1^5 m_3 + m_2^3 m_3(m_1^2 + m_2^2 
+ m_3^2)\right] -m_1^4m_2^2 I, \nonumber \\
&&L_2 = \frac{1}{16\pi^2}\mu^{2(n-3)}\left\{ m_2m_3 \left[\frac{1}{4}m_1^4 
+ m_2^4 + \frac{3}{4} m_1^2 m_2^2 - \frac{1}{4} m_1^2 m_3^2 + \frac{1}{3} m_2^2 m_3^2 \right]
-\frac{1}{4} m_1^3 m_2 \left[m_1^2 + m_2^2 - m_3^2 \right] \right.\nonumber \\
&&~~~~\left. +m_1m_3\left[\frac{1}{2} m_1^4 + \frac{1}{4} m_1^2(m_1^2 + m_2^2 - m_3^2)\right]\right\}-\left[ \frac{1}{4}m_1^6 + \frac{1}{2}m_1^4(m_2^2 
- m_3^2) + \frac{1}{4}m_1^2(m_2^2-m_3^2)^2\right] I, \nonumber \\
&& \tilde{L}_1 = \frac{15}{n(n-1)(n+2)(n+8)}[(3n+13)L_1 + 8(n-3)L_2], \nonumber \\
&& \tilde{L}_2 = \frac{15}{n(n-1)(n+2)(n+8)}[(n-3)L_1 + 2(5n-4)L_2],
 \nonumber \\
&&L_3 = \frac{3}{n(n-1)(n+2)(n+8)}[(3n+2)(n+7)L_1 + 4(n+4)(n-3)L_2], 
\nonumber \\
&&L_4 = \frac{15}{n(n+2)} L_1, 
~~~~L_5 = \frac{3}{n} L_1, \nonumber \\
&& L_6 = \frac{3}{n(n-1)(n+2)(n+8)}[(n+6)(n-3) L_1 + 2(3n^2+12n-8)L_2];
 \nonumber \\
&& R_1 = \frac{\mu^{2(n-3)}}{32\pi^2} [m_1^3 m_2^3 - m_3^3(m_1^3 
+ m_2^3) -m_3(m_1^5 + m_2^5)]  +  \frac{1}{2} (m_1^2 
+ m_2^2 - m_3^2) K_1,        \nonumber\\
&&R_2 = \frac{\mu^{2(n-3)}}{32\pi^2} [\frac{1}{3}(m_1^3 m_2^3 
- m_3^3 (m_1^3 + m_2^3)) -m_3(m_1^5 + m_2^5)] 
+ \frac{1}{2} (m_1^2 + m_2^2 - m_3^2) K_2,   \nonumber\\
&&R_3 = \frac{15}{n(n-1)(n+2)(n+4)}[R_1(5n-1) + 2 R_2(n-3)], \nonumber \\
&& R_4 = \frac{15}{n(n+2)} R_1, \nonumber \\
&& R_5 = \frac{3}{n} R_1. \nonumber
\end{eqnarray}
We have employed the following relations to facilitate
the two-loop calculation,
\begin{eqnarray}
&&\tilde{K}_1 - \tilde{K}_2 = \frac{6}{n(n-1)}(K_1 - K_2),\nonumber
\\
&&\tilde{L}_1 - \tilde{L}_2 = \frac{30}{n(n-1)(n+2)} (L_1-L_2), \nonumber \\
&&L_3-\tilde{L}_1 = \frac{3(n-3)}{n(n-1)(n+2)(n+8)}[(3n+17)L_1 + 4(n-6)L_2],
 \nonumber \\
&& L_4 - \tilde{L}_1 = \frac{3(n-3)}{n(n-1)(n+2)(n+8)}[(n+7)L_1 - 8L_2], \nonumber \\
&& L_5-L_4 = \frac{3(n-3)}{n(n+2)} L_1, \nonumber \\
&& L_6 - \tilde{L}_2 = \frac{3(n-3)}{n(n-1)(n+2)(n+8)}[(n+1)L_1 
+ 2(3n-4)L_2],\nonumber \\
&&R_4 - R_3 = \frac{15(n-3)}{n(n-1)(n+2)(n+4)}[(n+1)R_1 -2 R_2] \nonumber \\
&&R_5 - R_4 = \frac{3(n-3)}{n(n+2)}R_1. \nonumber
\end{eqnarray}

\begin{figure}[htb]
\begin{center}
\leavevmode
\epsfxsize=10 cm
\epsfbox{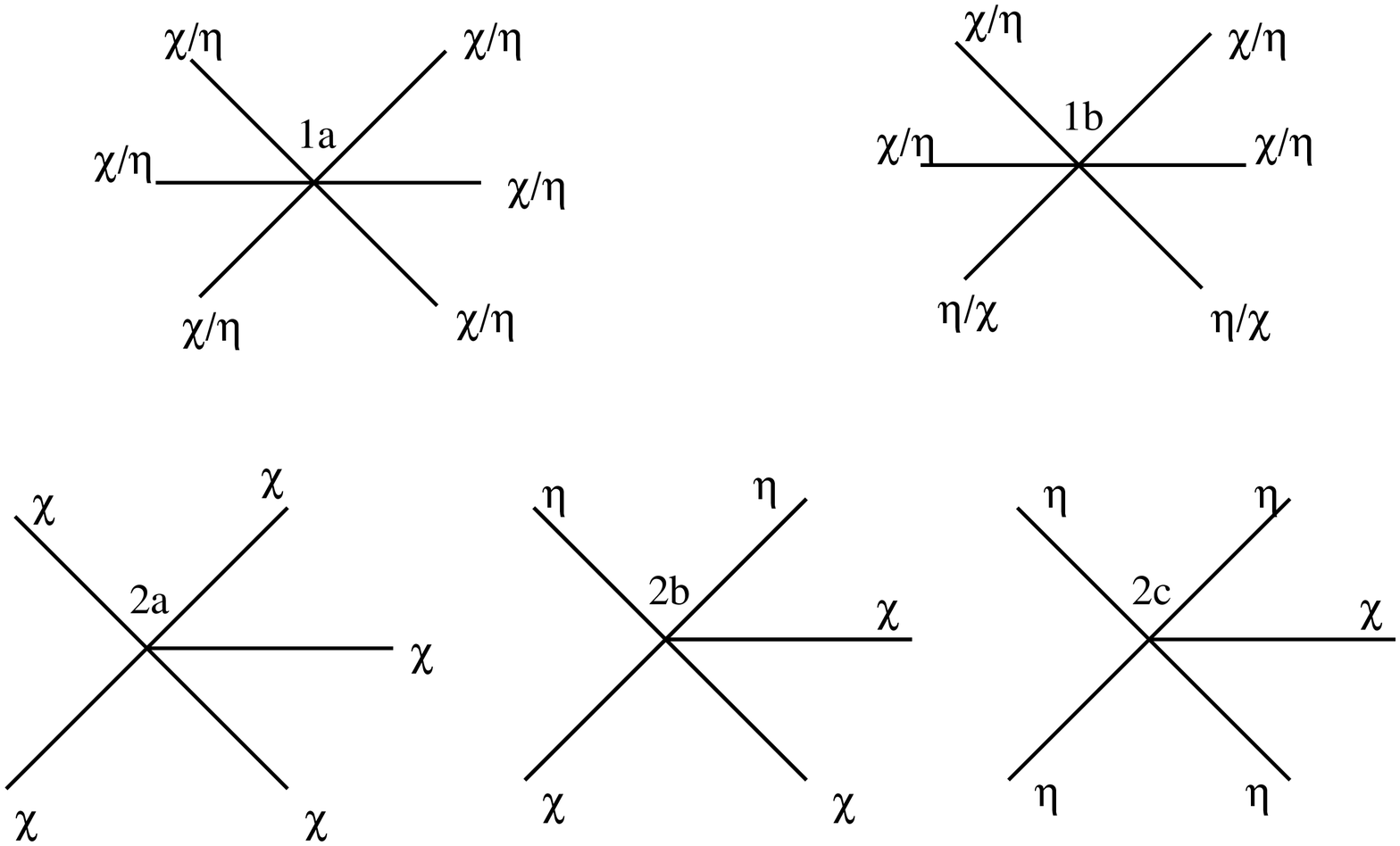}
\end{center}
\caption{Interaction vertices $V_{1a}$, $V_{1b}$,$V_{2a}$,$V_{2b}$,$V_{2c}$}
\label{ramp}
\end{figure}
\begin{figure}[htb]
\begin{center}
\leavevmode
\epsfxsize=10 cm
\epsfbox{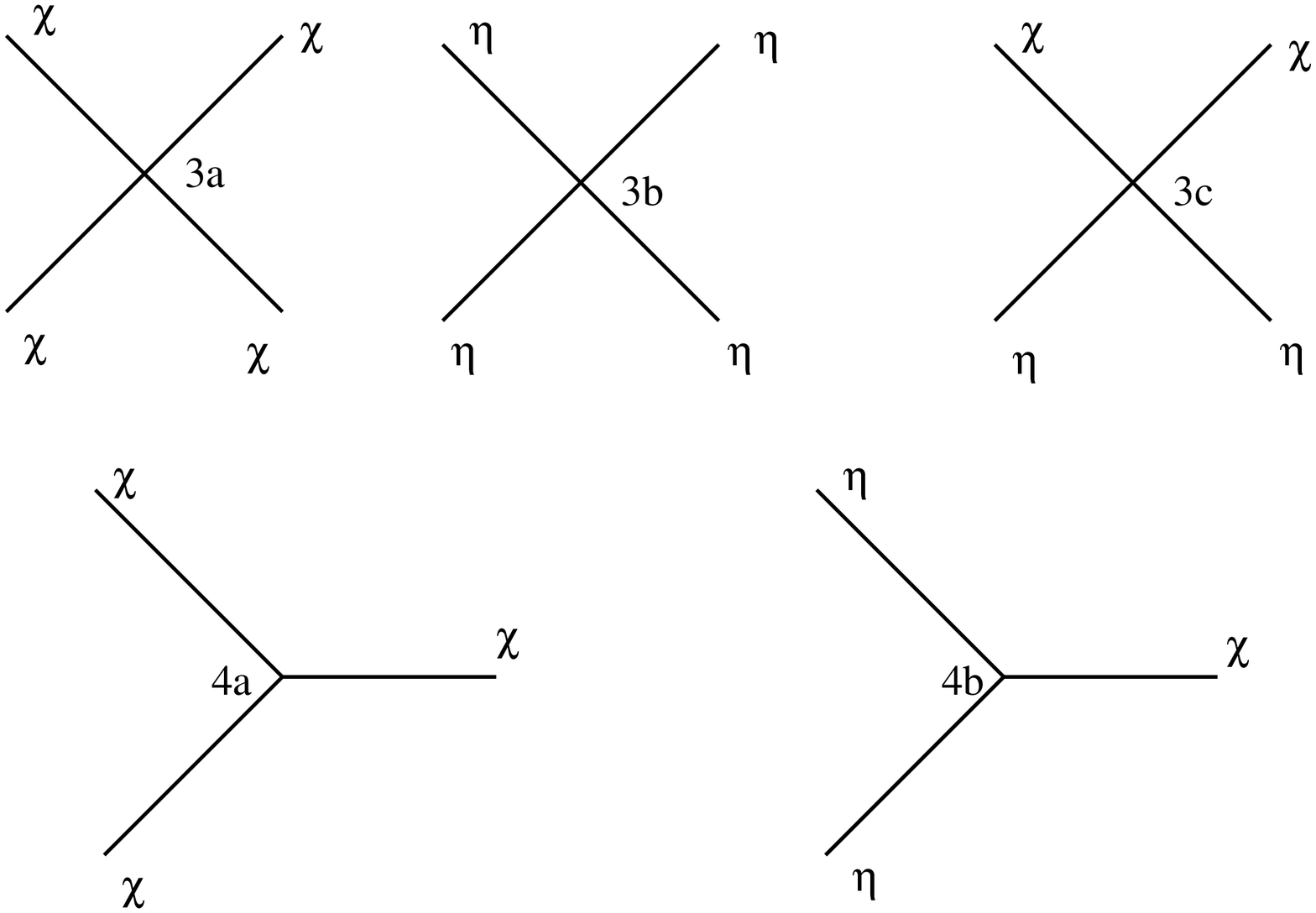}
\end{center}
\caption{Interaction vertices $V_{3a}$,$V_{3b}$,$V_{3c}$,$V_{4a}$,$V_{4b}$}
\label{ramp}
\end{figure}
 \begin{figure}[htb]
\begin{center}
\leavevmode
\epsfxsize=10 cm
\epsfbox{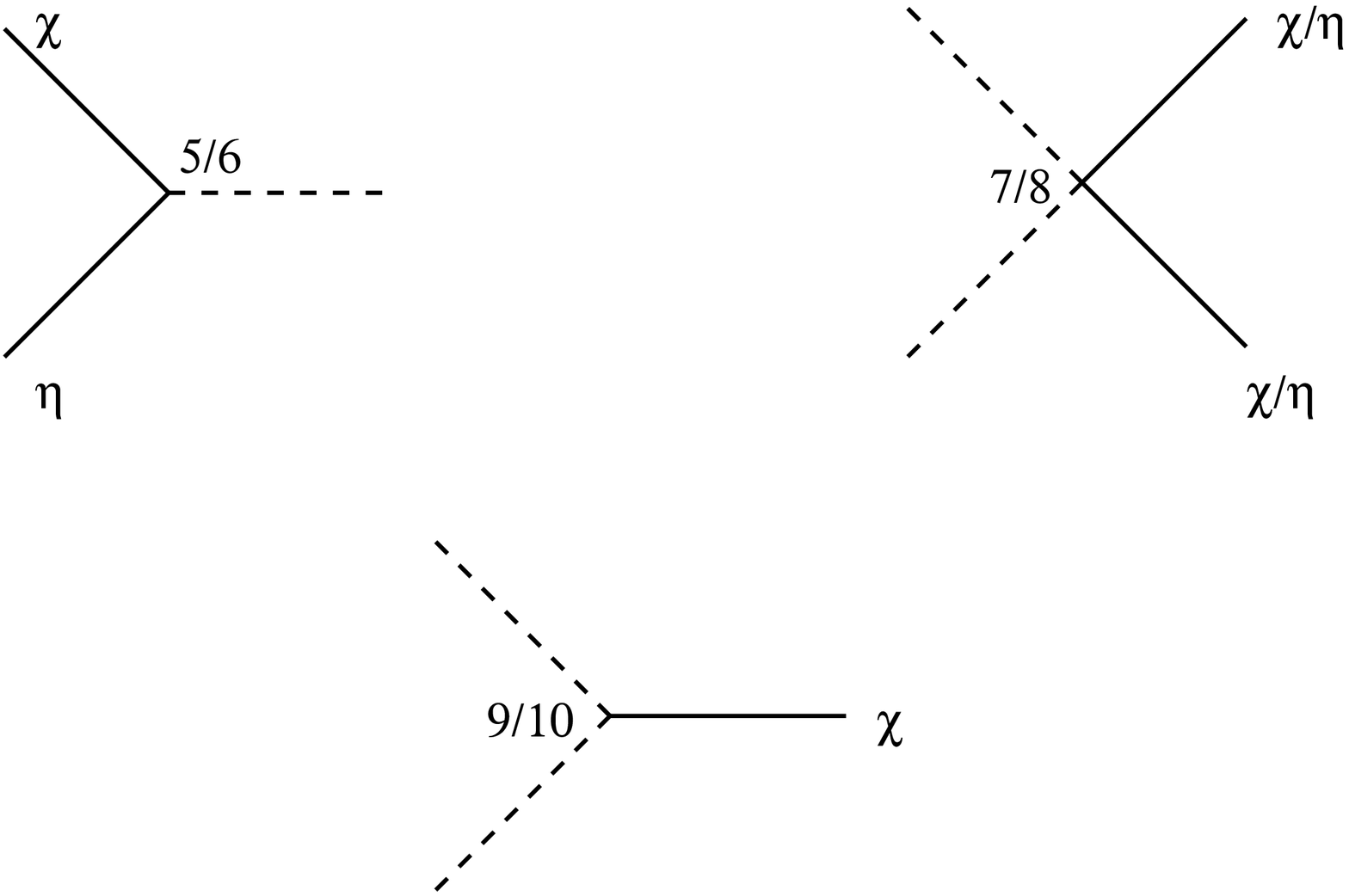}
\end{center}
\caption{Interaction vertices $V_5$,$V_6$,$V_7$,$V_8$,$V_9$,$V_{10}$}
\label{ramp}
\end{figure}
 \begin{figure}[htb]
\begin{center}
\leavevmode
\epsfxsize=10 cm
\epsfbox{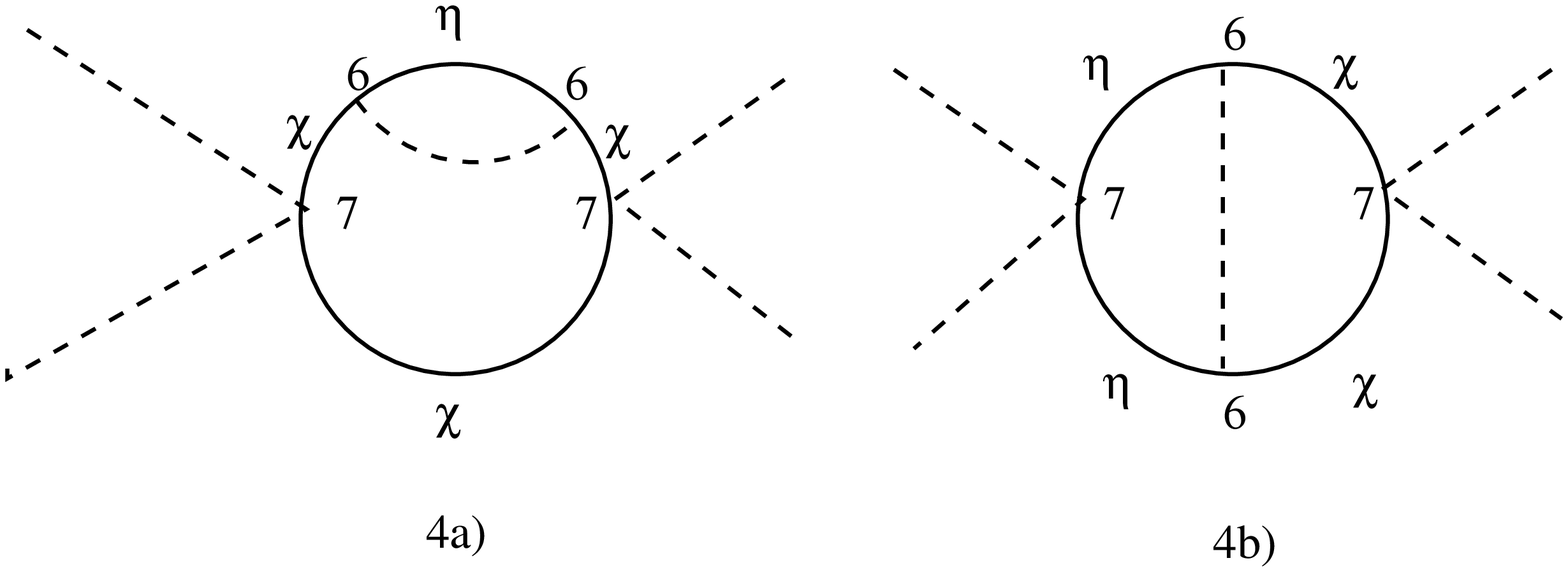}
\end{center}
\caption{Two loop contributions to $A^4$ part of the effective action.}
\label{ramp}
\end{figure}
 \begin{figure}[htb]
\begin{center}
\leavevmode
\epsfxsize=10 cm
\epsfbox{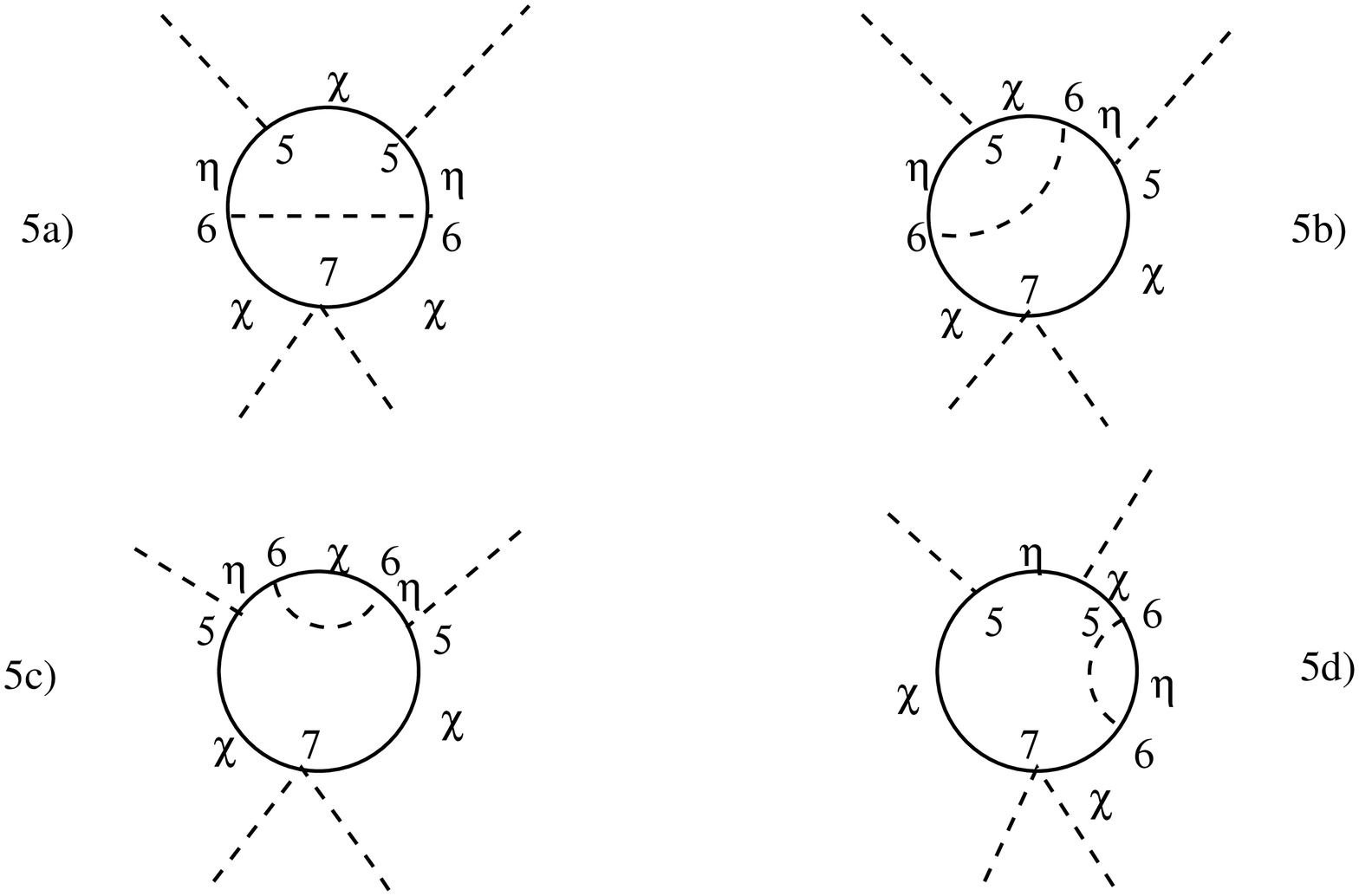}
\end{center}
\caption{Two loop contributions to $A^4$ part of the effective action.}
\label{ramp}
\end{figure}
 \begin{figure}[htb]
\begin{center}
\leavevmode
\epsfxsize=10 cm
\epsfbox{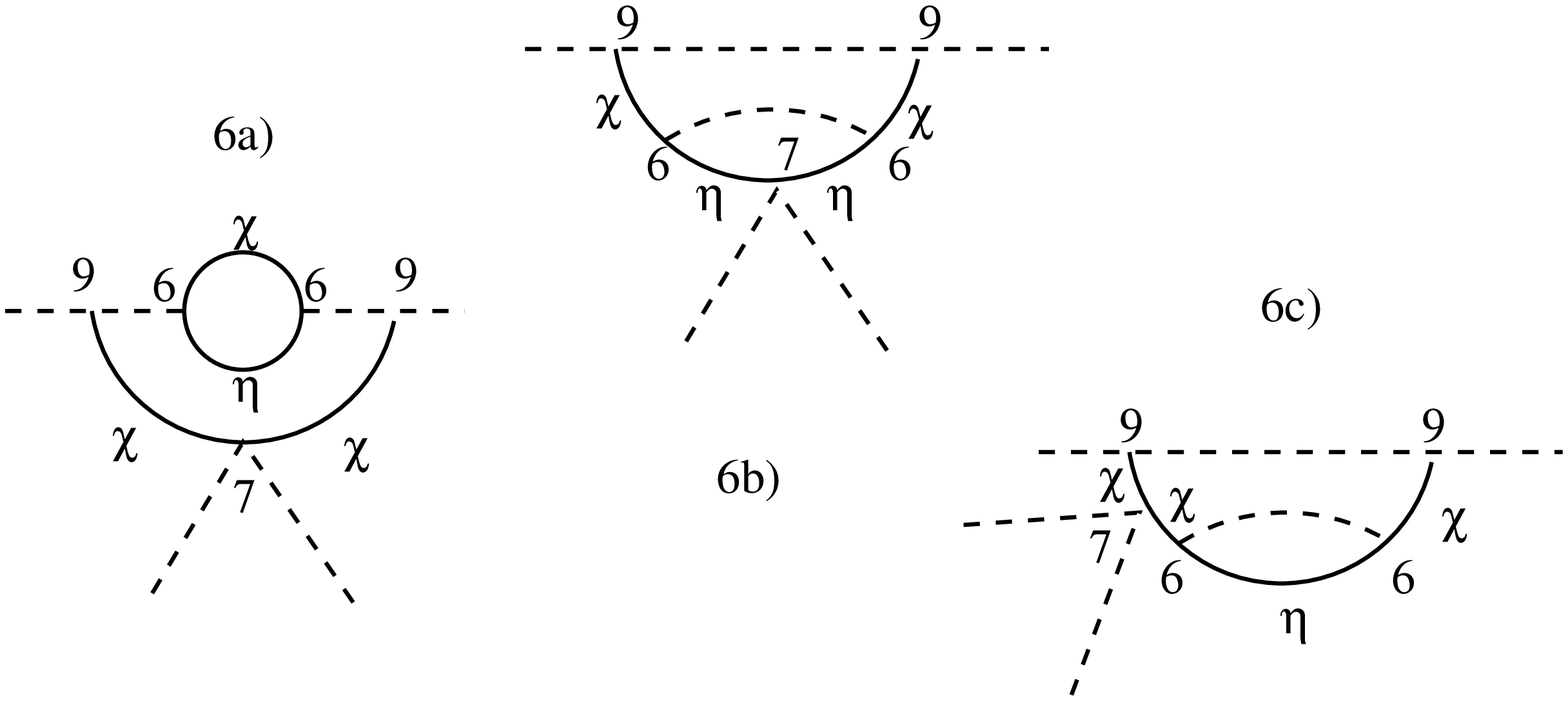}
\end{center}
\caption{Two loop contributions to $A^4$ part of the effective action.}
\label{ramp}
\end{figure}
 \begin{figure}[htb]
\begin{center}
\leavevmode
\epsfxsize=10 cm
\epsfbox{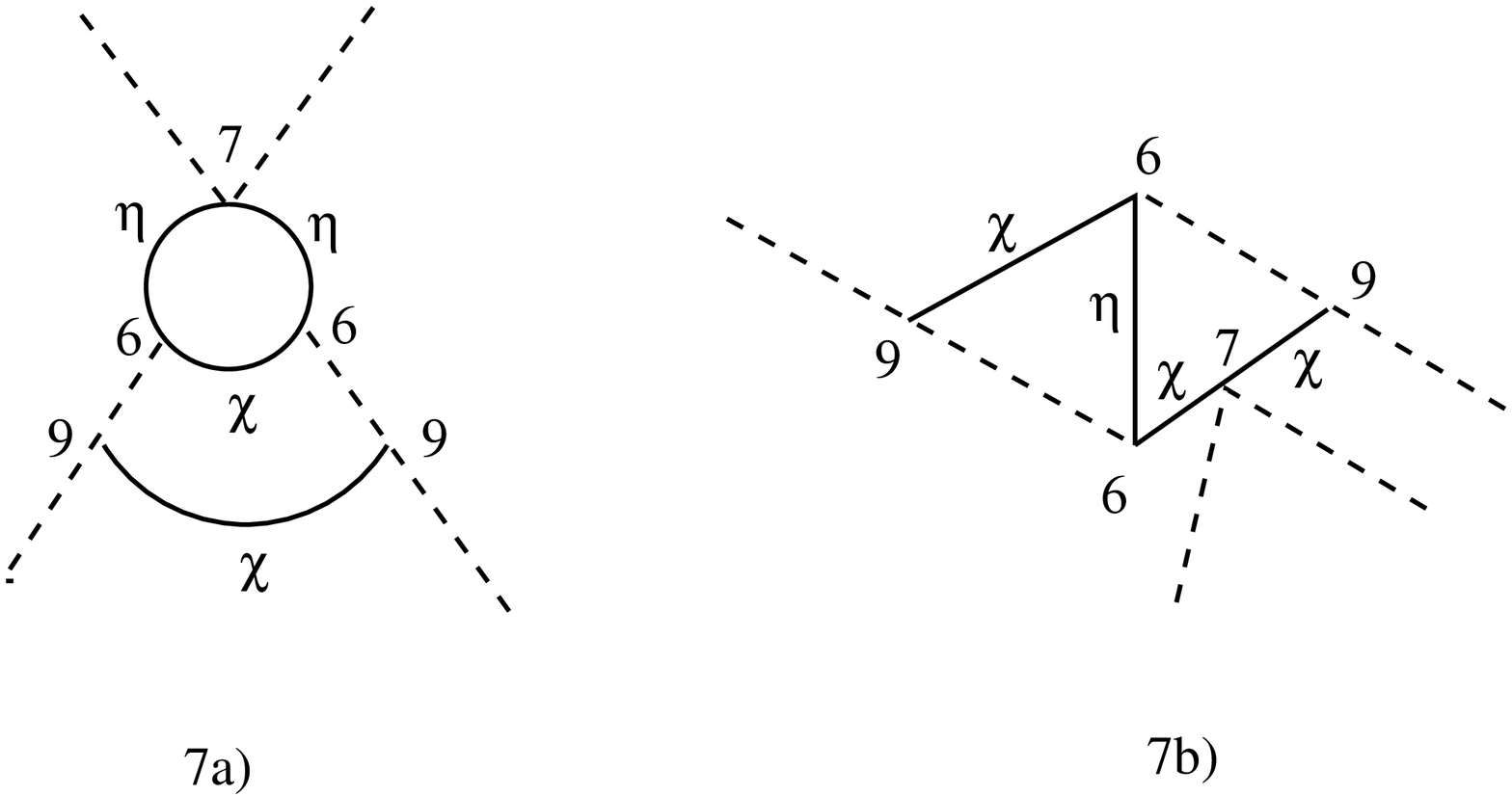}
\end{center}
\caption{Two loop contributions to $A^4$ part of the effective action.}
\label{ramp}
\end{figure}
\begin{figure}[htb]
\begin{center}
\leavevmode
\epsfxsize=10 cm
\epsfbox{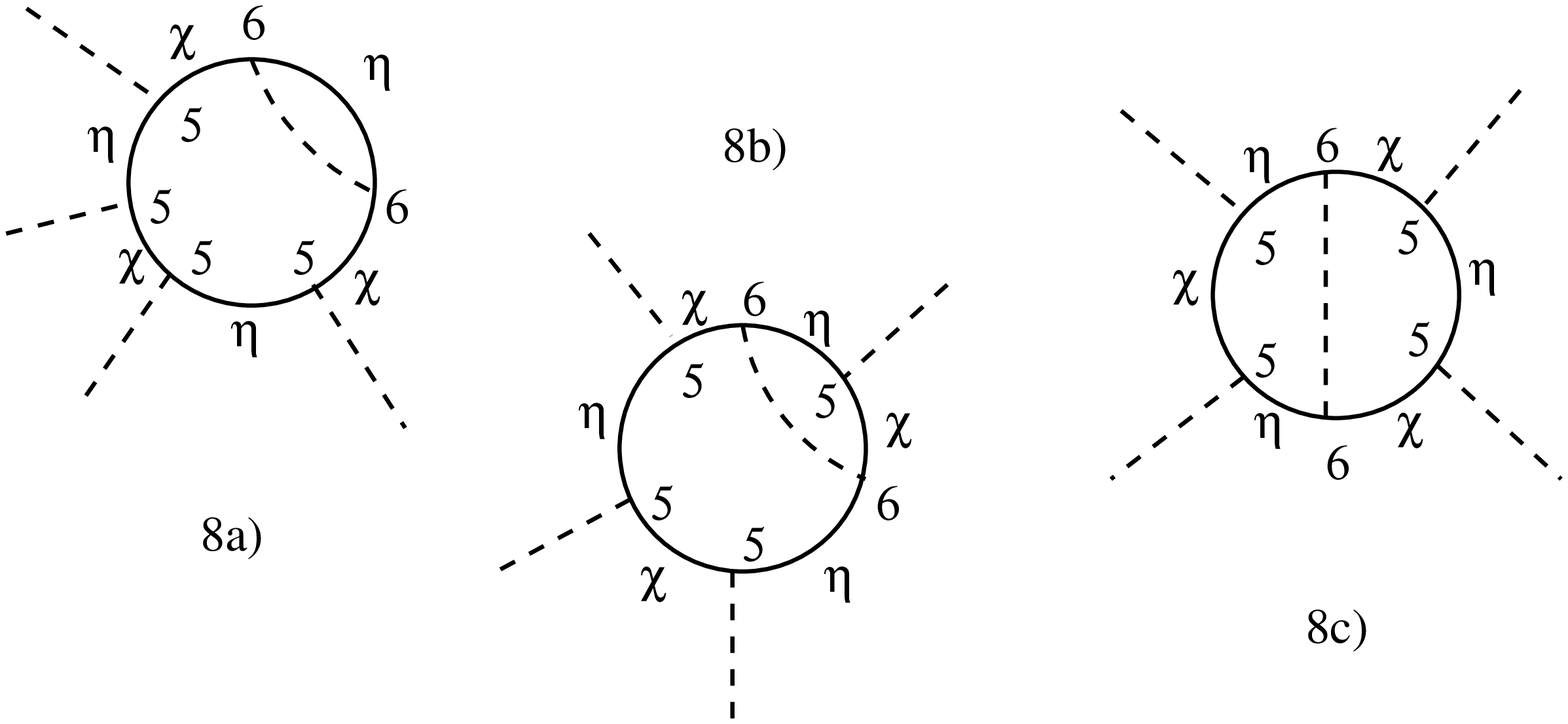}
\end{center}
\caption{Two loop contributions to $A^4$ part of the effective action.}
\label{ramp}
\end{figure}
 \begin{figure}[htb]
\begin{center}
\leavevmode
\epsfxsize=10 cm
\epsfbox{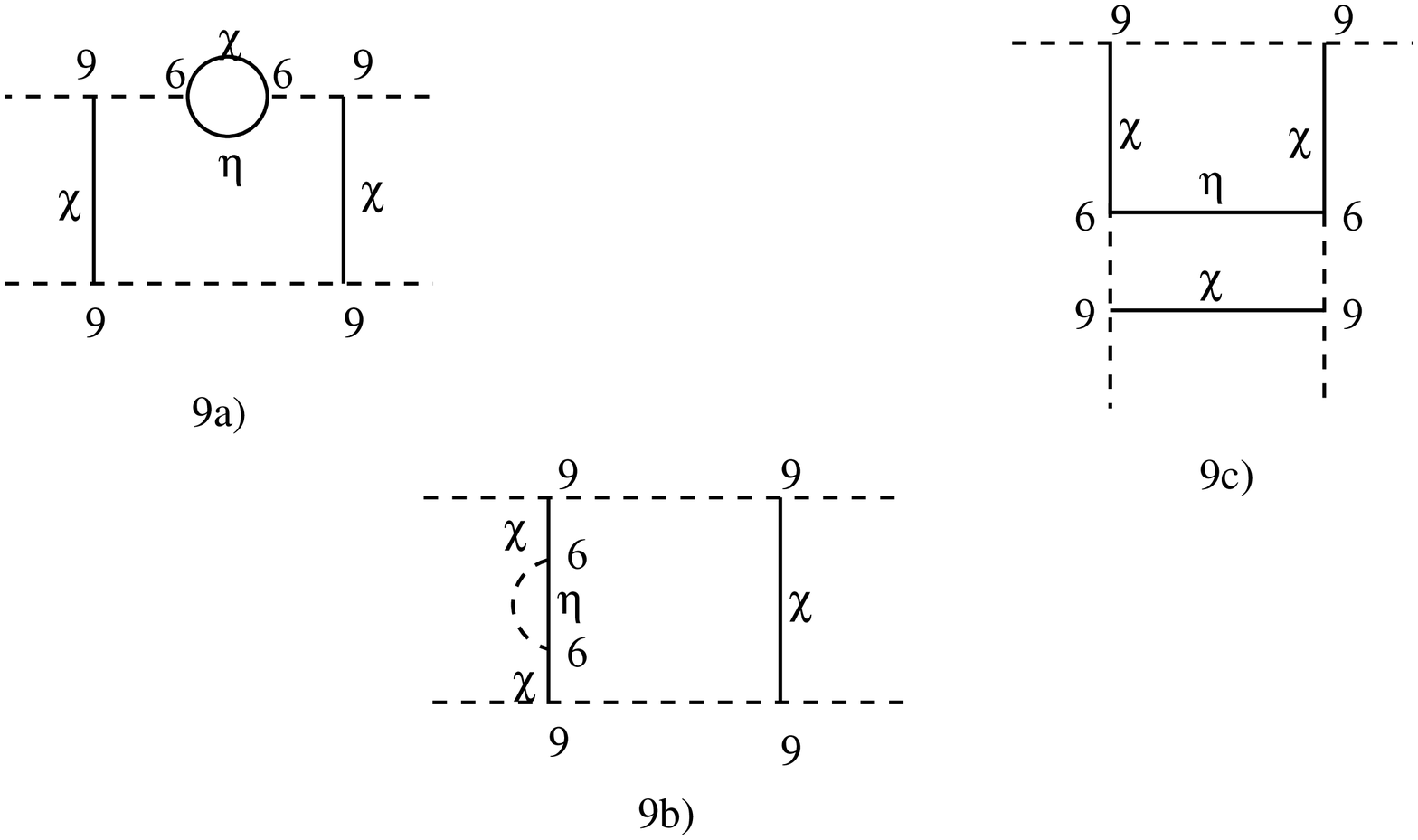}
\end{center}
\caption{Two loop contributions to $A^4$ part of the effective action.}
\label{ramp}
\end{figure}
\begin{figure}[htb]
\begin{center}
\leavevmode
\epsfxsize=10 cm
\epsfbox{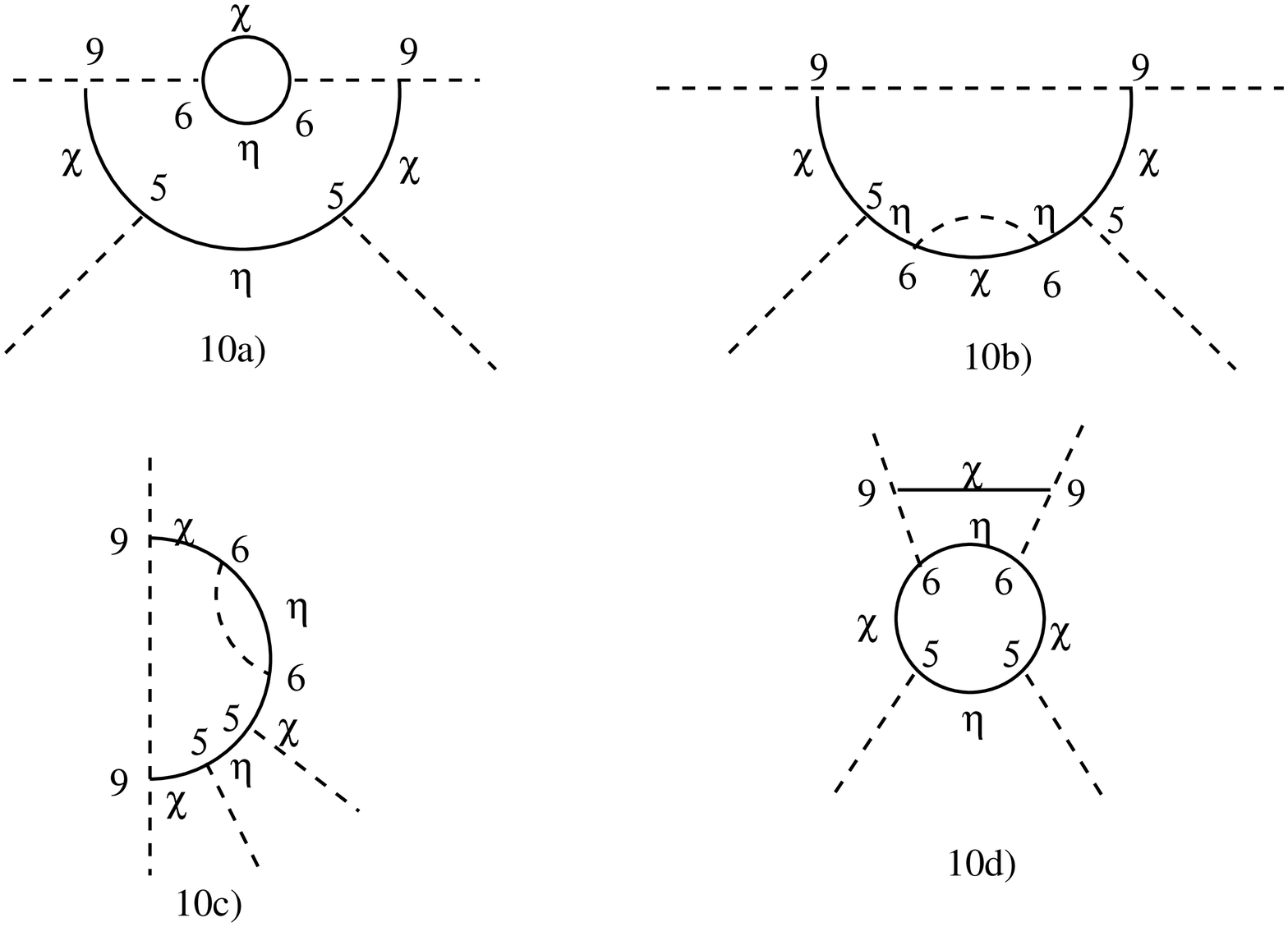}
\end{center}
\caption{Two loop contributions to $A^4$ part of the effective action.}
\label{ramp}
\end{figure}
\begin{figure}[htb]
\begin{center}
\leavevmode
\epsfxsize=10 cm
\epsfbox{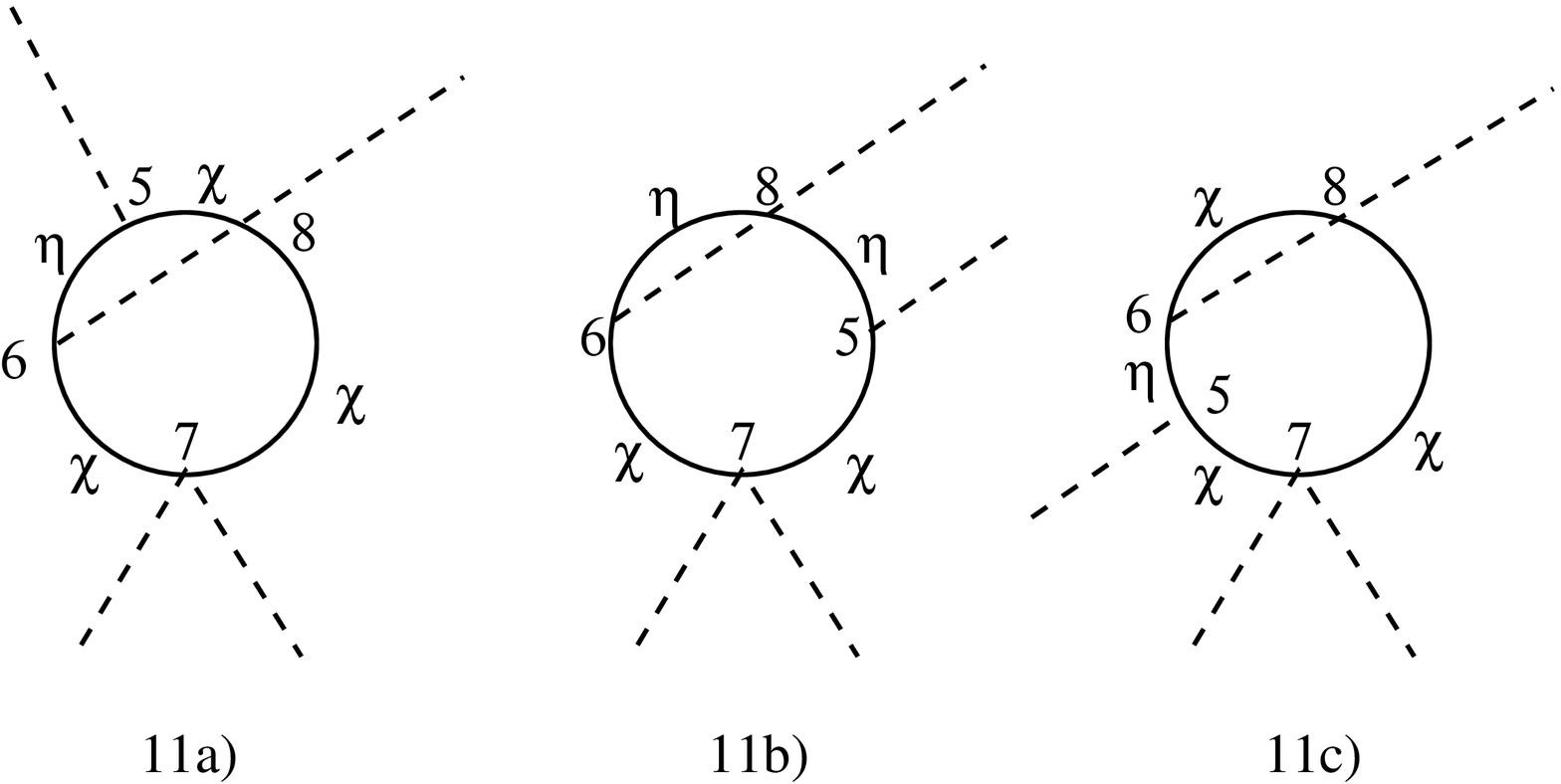}
\end{center}
\caption{Two loop contributions to $A^4$ part of the effective action.}
\label{ramp}
\end{figure}
\begin{figure}[htb]
\begin{center}
\leavevmode
\epsfxsize=10 cm
\epsfbox{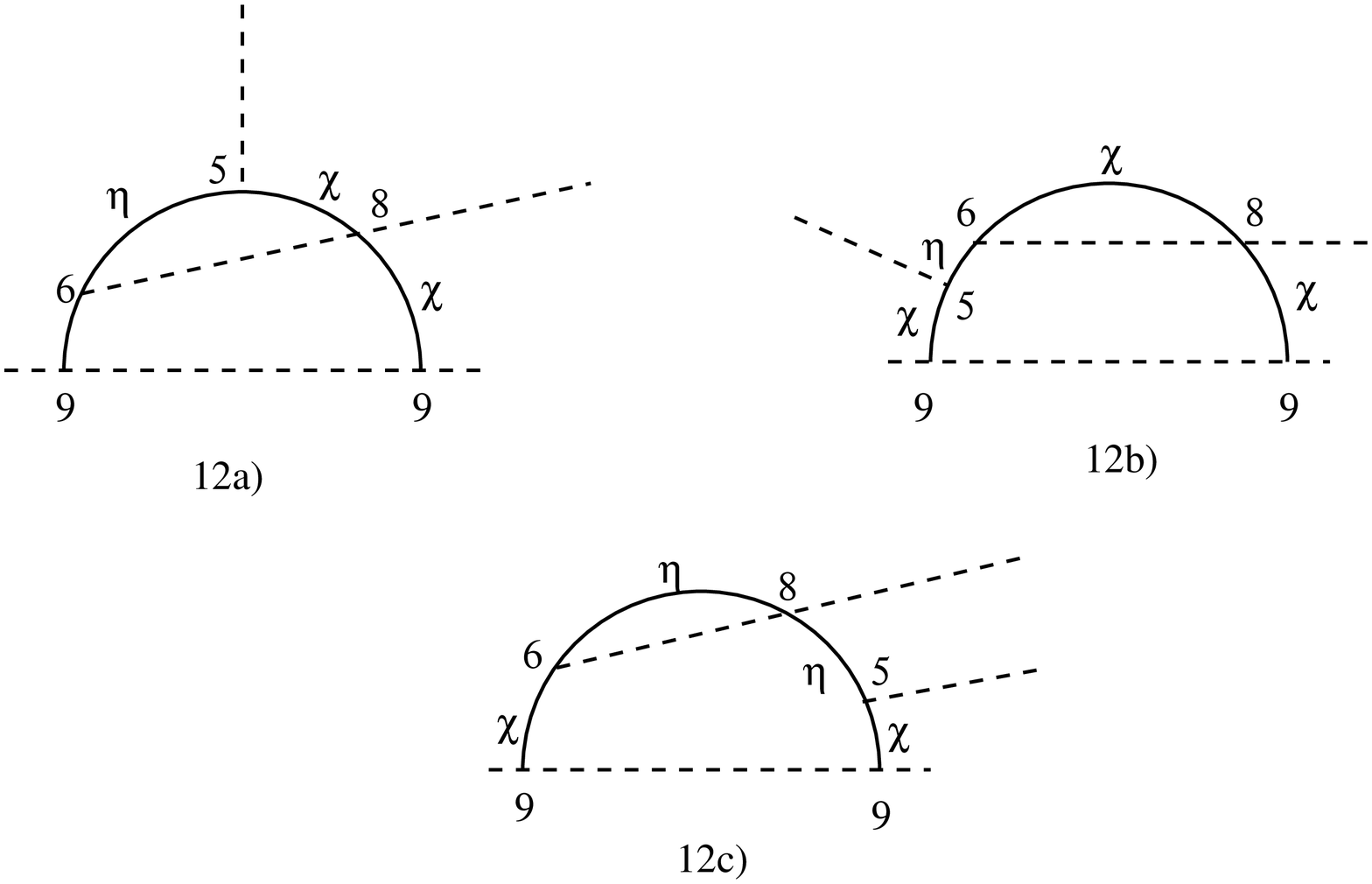}
\end{center}
\caption{Two loop contributions to $A^4$ part of the effective action.}
\label{ramp}
\end{figure}
\begin{figure}[htb]
\begin{center}
\leavevmode
\epsfxsize=10 cm
\epsfbox{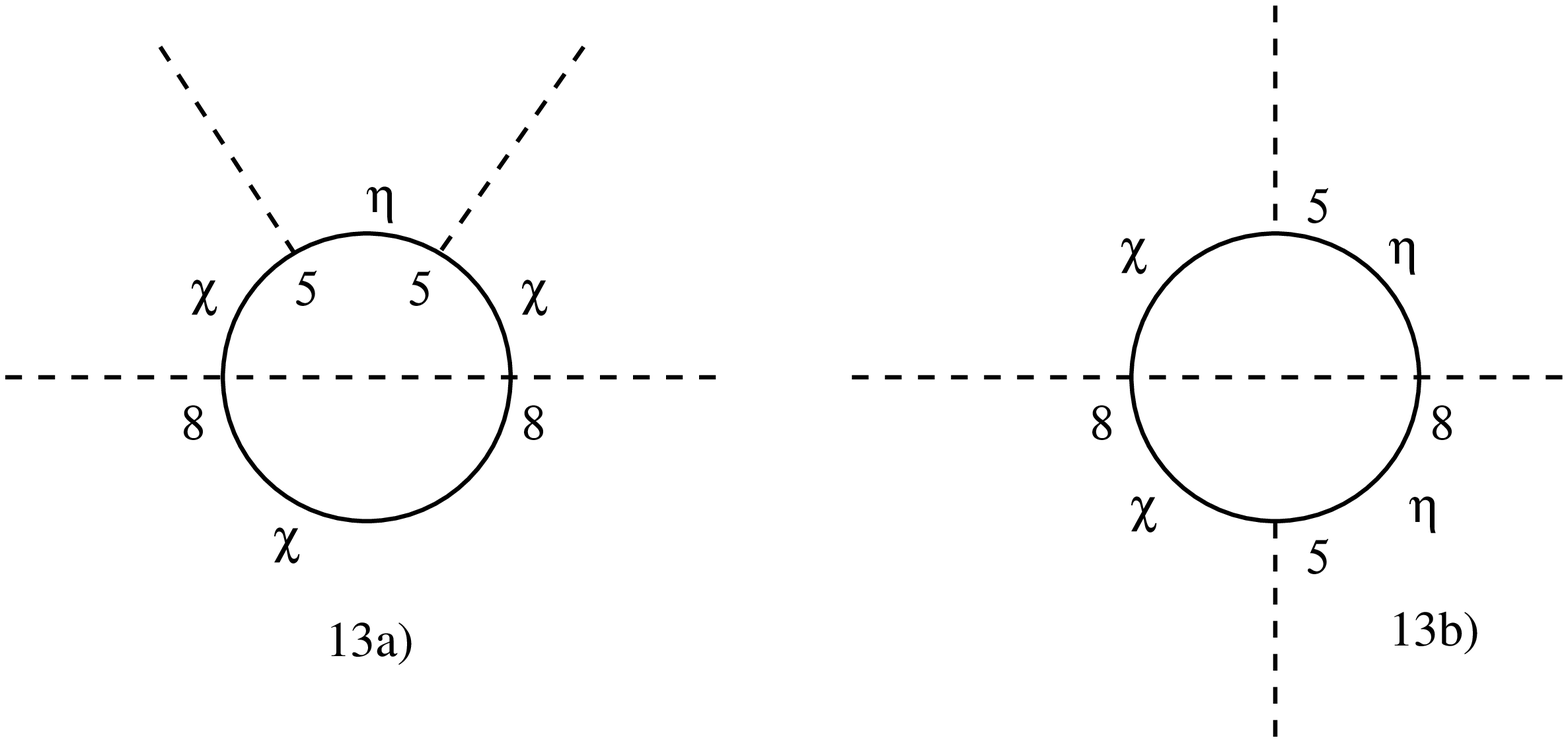}
\end{center}
\caption{Two loop contributions to $A^4$ part of the effective action.}
\label{ramp}
\end{figure}
\begin{figure}[htb]
\begin{center}
\leavevmode
\epsfxsize=10 cm
\epsfbox{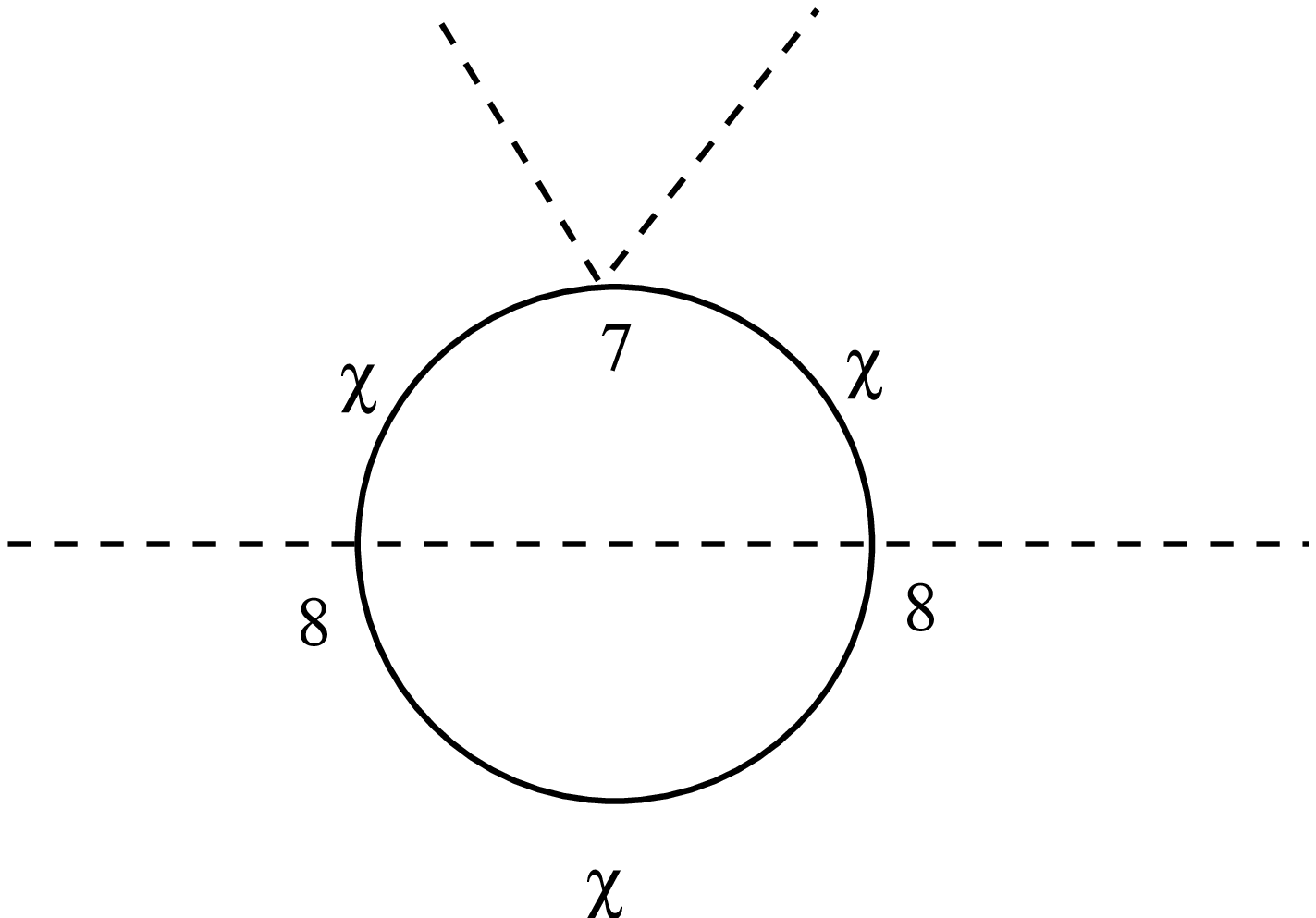}
\end{center}
\caption{Two loop contributions to $A^4$ part of the effective action.}
\label{ramp}
\end{figure}
\begin{figure}[htb]
\begin{center}
\leavevmode
\epsfxsize=10 cm
\epsfbox{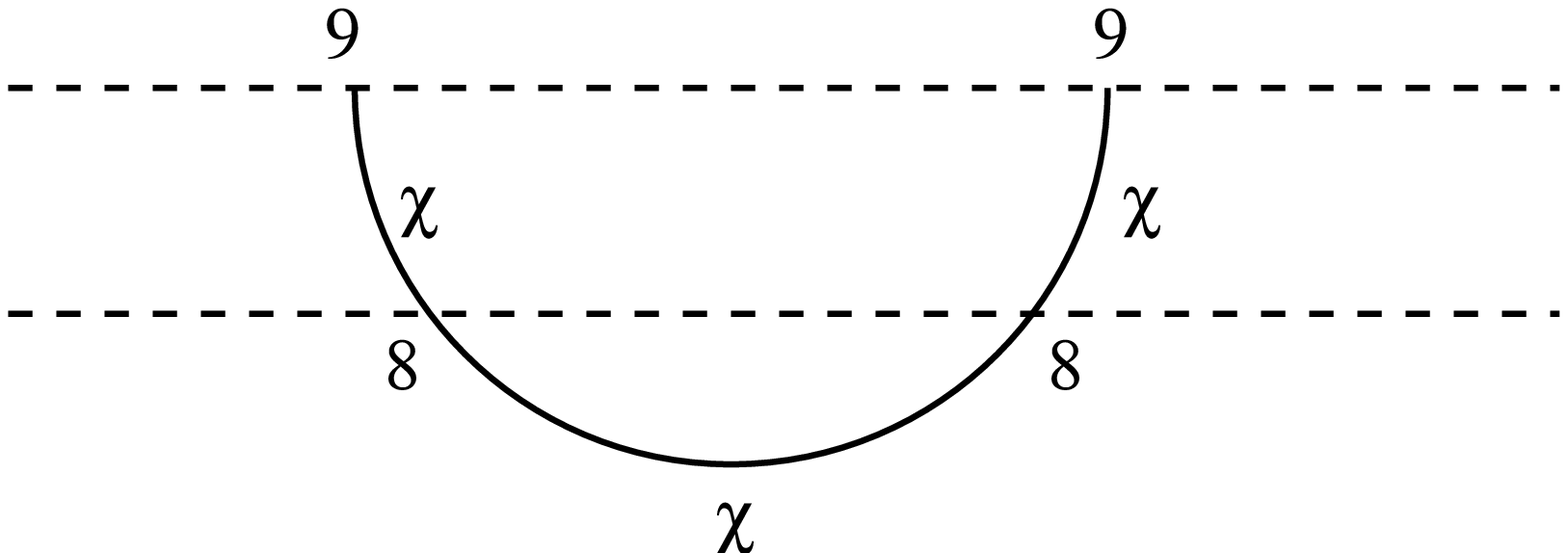}
\end{center}
\caption{Two loop contributions to $A^4$ part of the effective action.}
\label{ramp}
\end{figure}
\begin{figure}[htb]
\begin{center}
\leavevmode
\epsfxsize=10 cm
\epsfbox{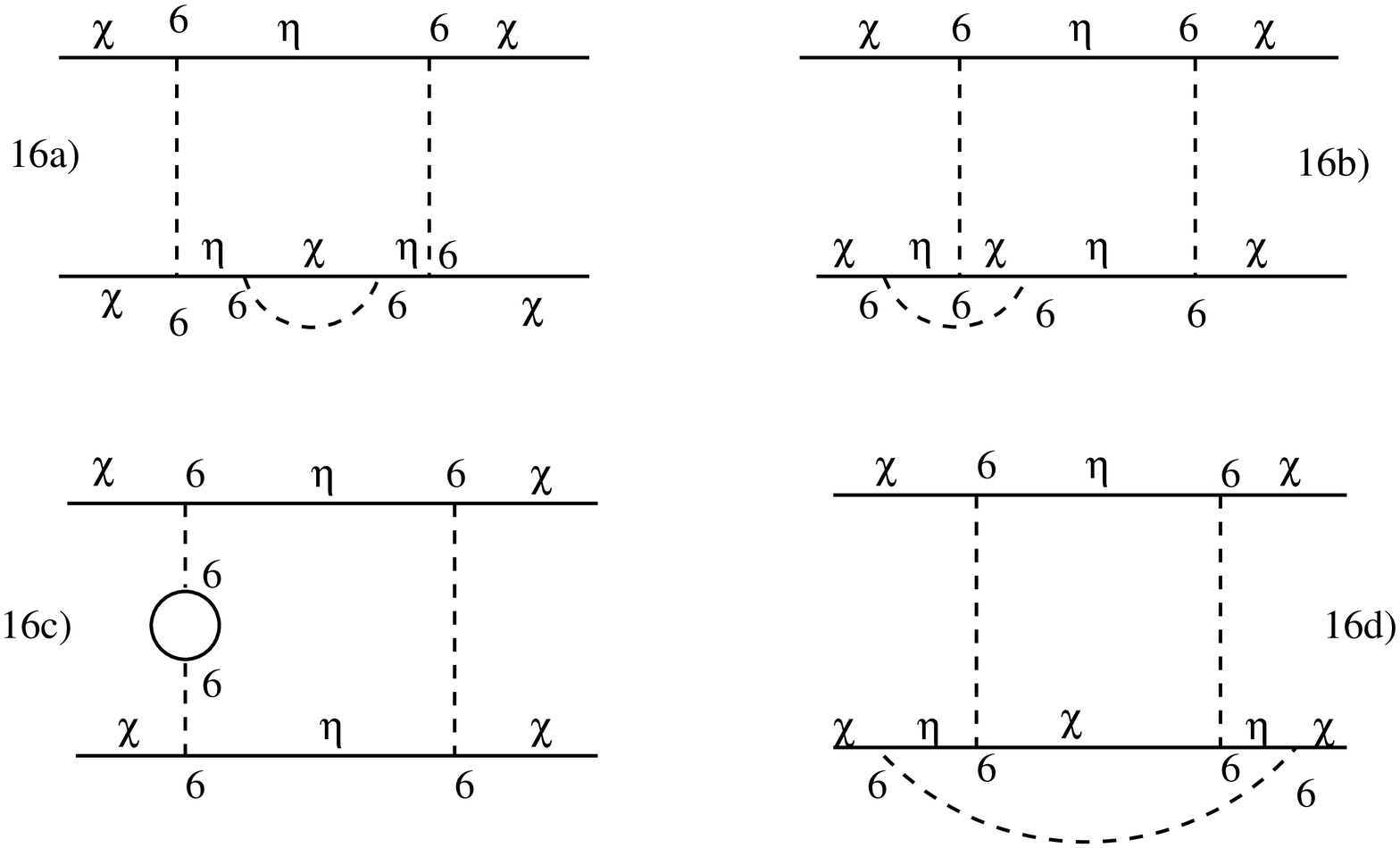}
\end{center}
\caption{Two loop contributions to $A^4$ part of the effective action.}
\label{ramp}
\end{figure}

\begin{figure}[htb]
\begin{center}
\leavevmode
\epsfxsize=10 cm
\epsfbox{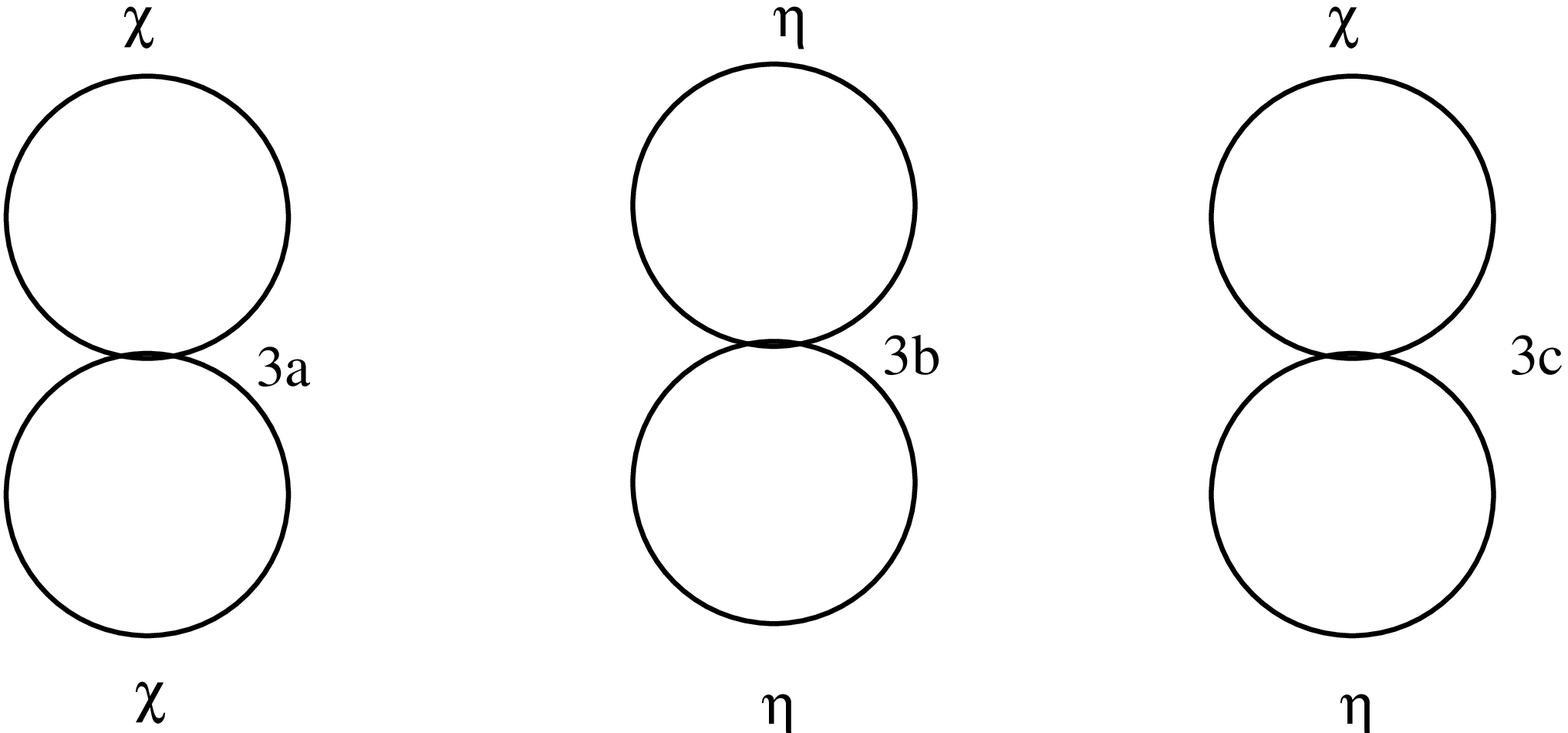}
\end{center}
\caption{Two loop contributions to the effective potential from two scalar 
loops.}
\label{ramp}
\end{figure}
\begin{figure}[htb]
\begin{center}
\leavevmode
\epsfxsize=10 cm
\epsfbox{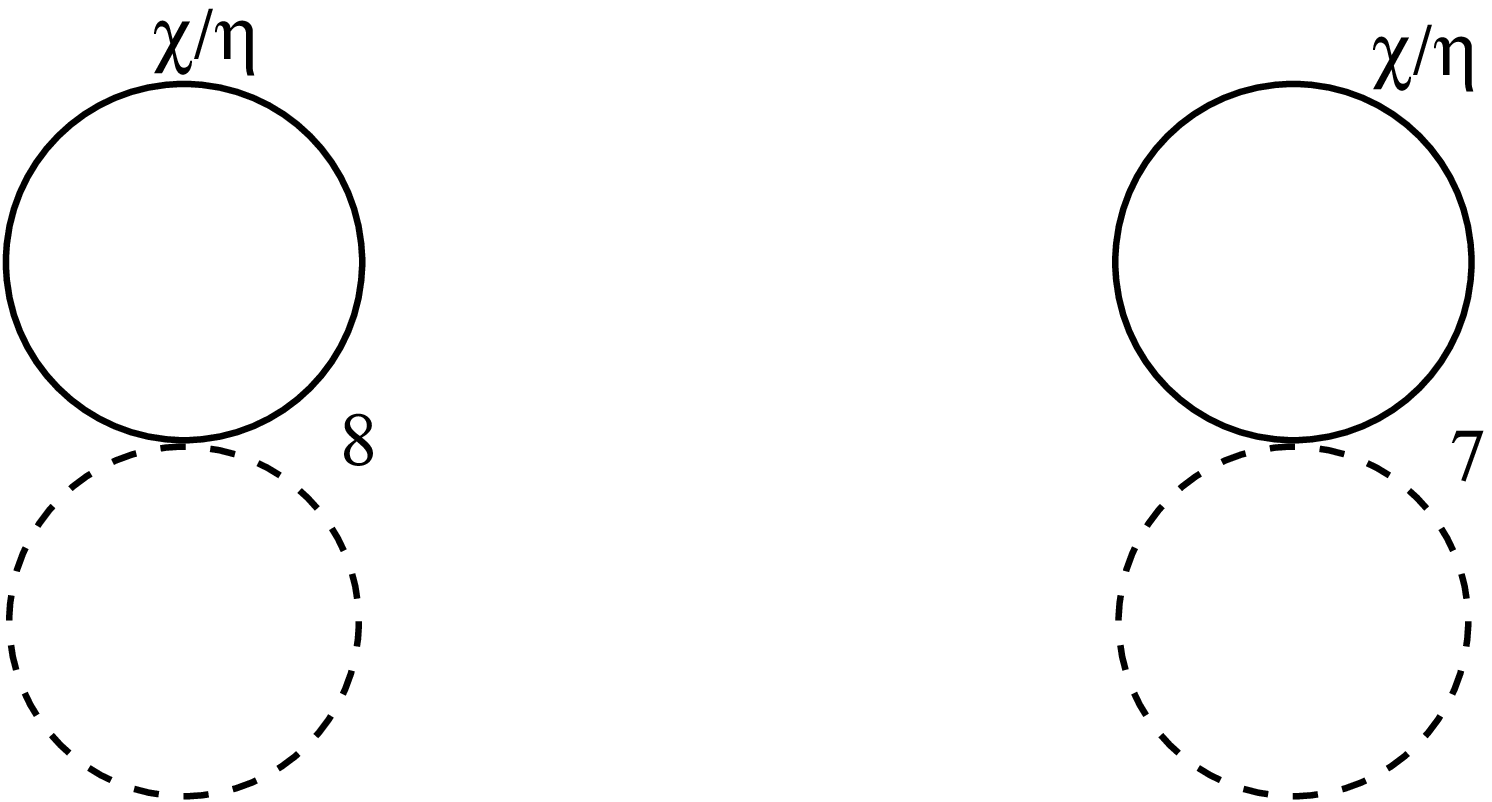}
\end{center}
\caption{Two loop contributions to the effective potential from one scalar and
one gauge field loop.}
\label{ramp}
\end{figure}
\begin{figure}[htb]
\begin{center}
\leavevmode
\epsfxsize=10 cm
\epsfbox{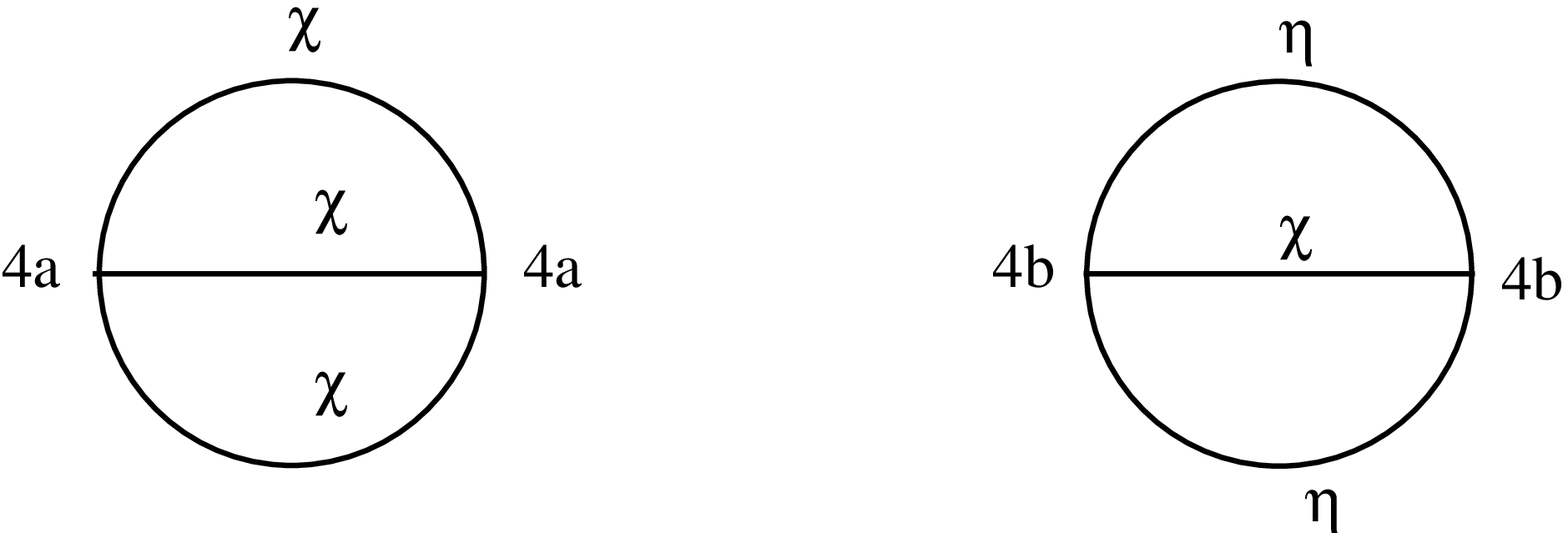}
\end{center}
\caption{Theta diagrams constructed from scalar propagators.}
\label{ramp}
\end{figure}
\begin{figure}[htb]
\begin{center}
\leavevmode
\epsfxsize=10 cm
\epsfbox{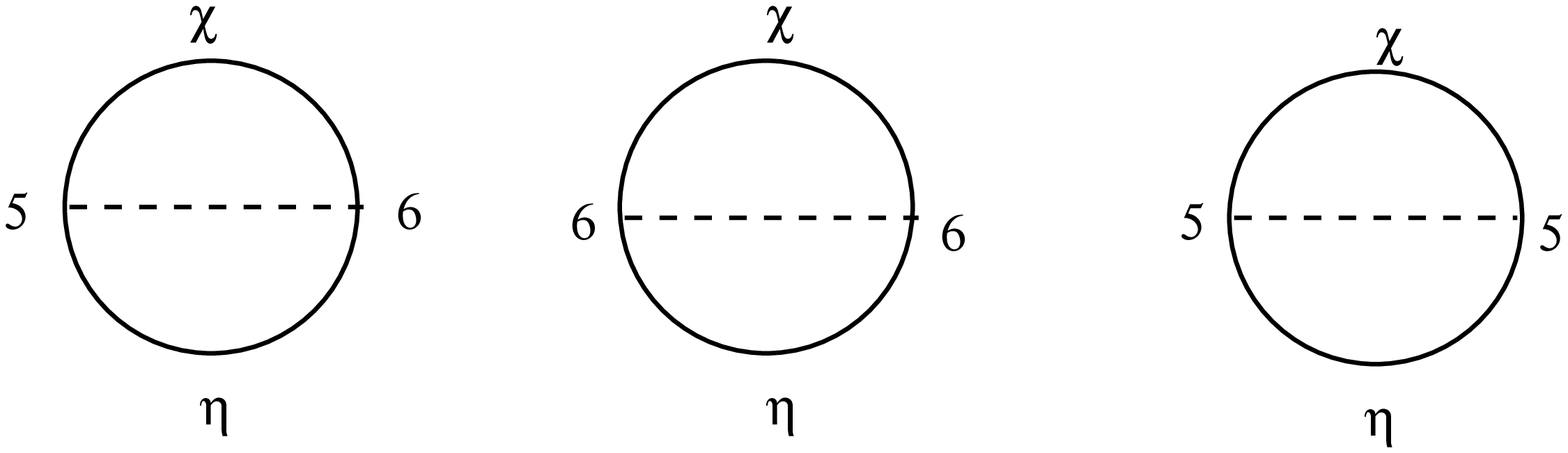}
\end{center}
\caption{Theta diagrams from one gauge field propagator and two scalar
propagators}
\label{ramp}
\end{figure}
\begin{figure}[htb]
\begin{center}
\leavevmode
\epsfxsize=10 cm
\epsfbox{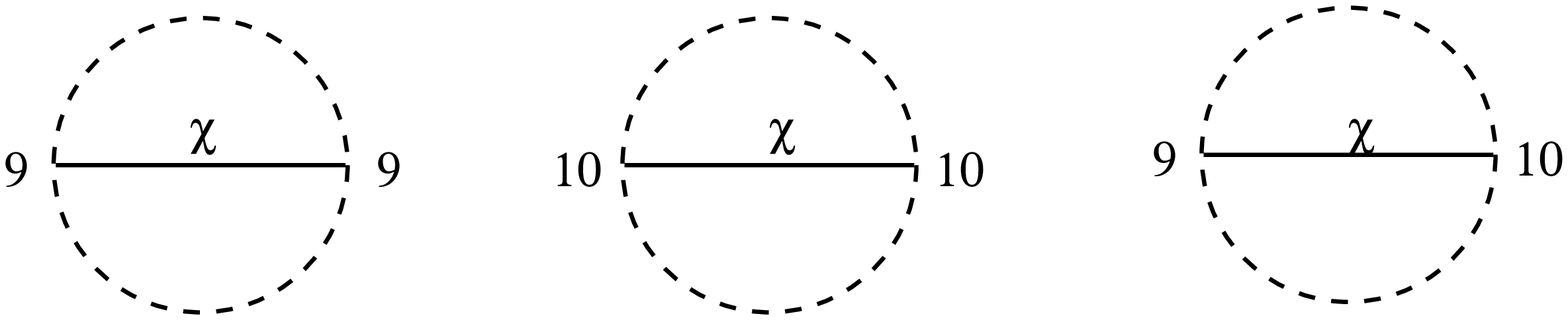}
\end{center}
\caption{Theta diagrams from two gauge field propagators and one scalar 
propagator.}
\label{ramp}
\end{figure}
\end{document}